\documentclass[acmsmall]{acmart}

\AtBeginDocument{%
  }

\setcopyright{rightsretained}
\acmDOI{10.1145/3660803}
\acmYear{2024}
\copyrightyear{2024}
\acmSubmissionID{fse24main-p734-p}
\acmJournal{PACMSE}
\acmVolume{1}
\acmNumber{FSE}
\acmArticle{96}
\acmMonth{7}
\received{2023-09-28}
\received[accepted]{2024-04-16}

\usepackage{algorithm}
\usepackage{algpseudocode}
\usepackage{graphicx}
\usepackage{textcomp}
\usepackage{xcolor}
\usepackage{balance}
\usepackage{xspace}
\usepackage{enumitem}
\usepackage{hyperref}
\usepackage[many]{tcolorbox}
\usepackage{subfigure}
\usepackage{booktabs}
\usepackage[normalem]{ulem}
\usepackage{colortbl}
\usepackage{siunitx}
\useunder{\uline}{\ul}{}
\usepackage[misc]{ifsym}

\newcommand{\issuesFull}{stereoscopic visual inconsistencies\xspace}
\newcommand{\issuesFullCap}{Stereoscopic Visual Inconsistencies\xspace}
\newcommand{\issues}{SVI issues\xspace}
\newcommand{\issueSingular}{SVI issue\xspace}
\newcommand{\issuesCap}{SVI Issues\xspace}
\newcommand{\issueCapSingular}{SVI Issue\xspace}
\newcommand{\issueAbbr}{SVI\xspace}

\newcommand{\tool}{\textsc{StereoID}\xspace}
\newcommand{\toolNoDepth}{\textsc{StereoID$_{ND}$}\xspace}
\newcommand{\toolFullUnderlined}{\textbf{\underline{ID}}entify \textbf{\underline{Stereo}}scopic visual inconsistencies\xspace}

\newcommand{\aemodelFullCap}{Depth-Aware Conditional Stereo Image Translator\xspace}
\newcommand{\aemodelFull}{depth-aware conditional stereo image translator\xspace}
\newcommand{\aemodelShortCap}{\textsc{Painter}\xspace}
\newcommand{\aemodelShort}{\textsc{Painter}\xspace}

\newcommand{\aemodelNoDepth}{\textsc{Painter$_{ND}$}\xspace}
\newcommand{\aemodelFullUnderlined}{de\textbf{\underline{P}}th-aw\textbf{\underline{A}}re cond\textbf{\underline{I}}tio\textbf{\underline{N}}al s\textbf{\underline{T}}ereo imag\textbf{\underline{E}} translato\textbf{\underline{R}}\xspace}
\newcommand{\aemodel}{\textsc{Painter}\xspace}

\newcommand{\numberOfPlatforms}{15\xspace}

\newcommand{\numberOfIssues}{282\xspace}
\newcommand{\numberOfIssuesWithScreenshots}{108\xspace}
\newcommand{\numberOfRealTestPairs}{82\xspace}

\newcommand{\numberOfAllScreenshotsManuallyCollected}{171,740\xspace}

\newtcolorbox{answerbox}{%
	left=3pt,
	right=3pt,
	top=0pt,
	bottom=0pt,
	boxrule=0mm,
	colback=black!10!white,
	breakable,
	frame empty
}%

\theoremstyle{definition}

\newcommand{\mytodocomment}[2]{}

\definecolor{applegreen}{rgb}{0.55, 0.80, 0.4}

\let\diff\undefined %

\ifdefined\diff
\newcommand{\textred}[1]{\textcolor{red}{#1}}
\newcommand{\textblue}[1]{\textcolor{blue}{#1}}
\newcommand{\textgreen}[1]{\textcolor{applegreen}{#1}}
\else
\newcommand{\textred}[1]{#1}
\newcommand{\textblue}[1]{#1}
\newcommand{\textgreen}[1]{#1}
\fi

\definecolor{mygray}{gray}{.9}

\begin{document}

\title[Less Cybersickness, Please: Demystifying and Detecting \issuesFullCap in VR Apps]{Less Cybersickness, Please: Demystifying and Detecting \issuesFullCap in Virtual Reality Apps}

\renewcommand*{\thefootnote}{\fnsymbol{footnote}}

\author{Shuqing Li}
\orcid{0000-0001-6323-1402}
\affiliation{%
  \institution{The Chinese University of Hong Kong}
  \city{Hong Kong}
  \country{China}
}
\email{sqli21@cse.cuhk.edu.hk}

\author{Cuiyun Gao}
\authornote{Corresponding author.}
\orcid{0000-0003-4774-2434}
\affiliation{%
  \institution{Harbin Institute of Technology}
  \city{Shenzhen}
  \country{China}
}
\email{gaocuiyun@hit.edu.cn}

\author{Jianping Zhang}
\orcid{0000-0002-8262-9608}
\affiliation{%
  \institution{The Chinese University of Hong Kong}
  \city{Hong Kong}
  \country{China}
}
\email{jpzhang@cse.cuhk.edu.hk}

\author{Yujia Zhang}
\orcid{0009-0001-3869-6344}
\affiliation{%
  \institution{Harbin Institute of Technology}
  \city{Shenzhen}
  \country{China}
}
\email{200110910@stu.hit.edu.cn}

\author{Yepang Liu}
\authornote{Yepang Liu is affiliated with both the Research Institute of Trustworthy Autonomous Systems and the Department of Computer Science and Engineering at Southern University of Science and Technology.}
\orcid{0000-0001-8147-8126}
\affiliation{%
  \institution{Southern University of Science and Technology}
  \city{Shenzhen}
  \country{China}
}
\email{liuyp1@sustech.edu.cn}

\author{Jiazhen Gu}
\orcid{0000-0002-5831-9474}
\affiliation{%
  \institution{The Chinese University of Hong Kong}
  \city{Hong Kong}
  \country{China}
}
\email{jiazhengu@cuhk.edu.hk}

\author{Yun Peng}
\orcid{0000-0003-1936-5598}
\affiliation{%
  \institution{The Chinese University of Hong Kong}
  \city{Hong Kong}
  \country{China}
}
\email{ypeng@cse.cuhk.edu.hk}

\author{Michael R. Lyu}
\orcid{0000-0002-3666-5798}
\affiliation{%
  \institution{The Chinese University of Hong Kong}
  \city{Hong Kong}
  \country{China}
}
\email{lyu@cse.cuhk.edu.hk}

\renewcommand*{\thefootnote}{\arabic{footnote}}
\setcounter{footnote}{0}

\begin{abstract}
The quality of Virtual Reality (VR) apps is vital, particularly the rendering quality of the VR Graphical User Interface (GUI). Different from traditional two-dimensional (2D) apps, VR apps create a 3D digital scene
for users, by rendering two distinct 2D images
for the user’s left and right eyes, respectively.
Stereoscopic visual inconsistency (denoted as ``SVI'') issues, however, undermine the rendering process of the user's brain,
leading to user discomfort and even adverse health effects.
Such issues commonly exist in VR apps but remain under-explored. To comprehensively understand the \issues, we conduct an empirical analysis on
282 \issueAbbr bug reports collected from 15 VR platforms, summarizing 15 types of manifestations of the issues.
The empirical analysis reveals that automatically detecting \issues is challenging, mainly because:
(1) lack of training data;
(2) the manifestations of \issues are diverse, complicated, and often application-specific; 
(3) most accessible VR apps are closed-source commercial software, we have no access to code, scene configurations, etc. for issue detection.
Our findings imply that the existing pattern-based supervised classification approaches may be inapplicable or ineffective in detecting the \issues.

To counter these challenges, we propose an 
unsupervised black-box testing framework named \tool to identify the \issuesFull, based only on the rendered GUI states.
\tool generates a synthetic right-eye image based on the actual left-eye image and computes distances between the synthetic right-eye image and the actual right-eye image to detect \issues.
We propose a \aemodelFull to power the image generation process,
which captures the expected perspective shifts 
between left-eye and right-eye images.
We build a large-scale unlabeled VR stereo screenshot dataset with larger than 171K images from 288 real-world VR apps, which can be utilized to train our \aemodelFull and evaluate the whole testing framework \tool.
After substantial experiments, \aemodelFull demonstrates superior performance in generating stereo images, outpacing traditional architectures. It achieved the lowest average L1 and L2 losses and the highest SSIM score, signifying its effectiveness in pixel-level accuracy and structural consistency for VR apps.
\tool further demonstrates its power for detecting \issues in both user reports and wild VR apps.
In summary, this novel framework enables effective
detection of elusive SVI issues, benefiting
the quality of VR apps.
\end{abstract}



 \maketitle

\section{Introduction}

\begin{figure}[t!] 
	\centering 
	\includegraphics[width=0.3\columnwidth]{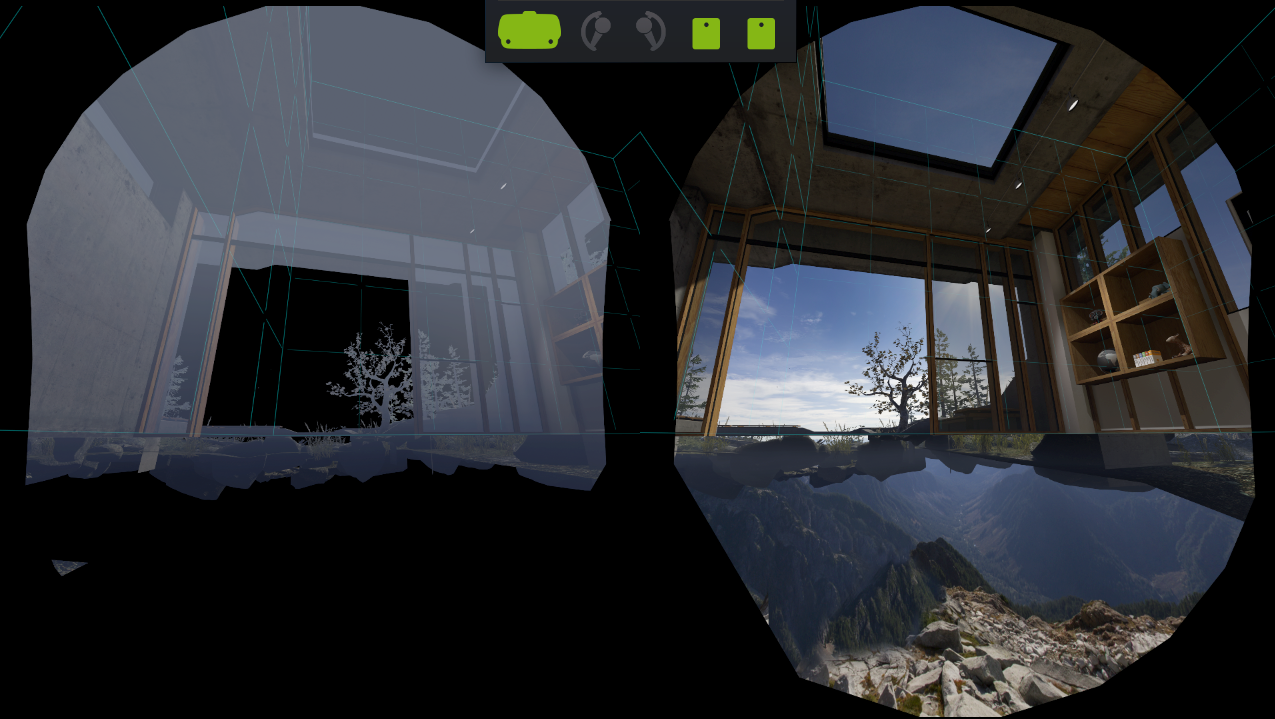} 
 \vspace{-0.5em}
	\caption{A bug causing \issuesFull in the SteamVR home app (issue \#299 of ValveSoftware/SteamVR-for-Linux~\protect\cite{website:github-steamvr-linux} on GitHub)}
	\label{fig:intro-issue-example-steamvr-home}
  \vspace{-1em}
\end{figure}

Virtual Reality (VR) is a technology that provides users with immersive experiences by creating interactive virtual environments.
Over the past few years, VR has experienced a remarkable surge in popularity and diversity,
encompassing tens of thousands of apps~\cite{paper:vr-software-quality} tailored for various purposes such as skill training~\cite{website:xr-application-VirtualSkill}, entertainment~\cite{website:xr-application-xrgames, website:xr-application-xrfilm}, and even usage scenarios that require high reliability like medical procedures~\cite{paper:vr-ar-in-surgery}. This successful deployment has captivated a global user base of exceeding 171 million people~\cite{website:vr-users-number}. %
VR apps adopt stereoscopic 3D (S3D)~\cite{paper:s3d-1, paper:s3d-2}, which provides two distinct two-dimensional (2D) images for the eyes of the user respectively. The user's brain then constructs the corresponding stereoscopic 3D scene with an illusion of depth based on these two images. 
Rendering issues can lead to discomfort feelings in VR, which is the well-known cybersickness\footnote{We use cybersickness and VR sickness interchangeably in this paper.}~\cite{paper:cybersickness-name-1, paper:vr-sickness-cause-measure-3} problem.
The cybersickness, including symptoms such as headaches, disorientation, and nausea, potentially affects
the users' health and safety, hindering the development and growth of VR apps~\cite{paper:issre-webxr-empirical,paper:vr-sickness-15-mins,paper:sigchi-2020-vr-sickness-symptoms-2, paper:walkingvibe-chi20,paper:nichols2002-vr-health-4}. 
A common cause of rendering-induced cybersickness issues is the inconsistent left and right eye view from the 2D-to-S3D construction process. Figure~\ref{fig:intro-issue-example-steamvr-home} illustrates an example.
This issue reported on GitHub~\cite{website:github-steamvr-linux} pertains to left-eye rendering corruption in the entry-point application for all Steam VR apps, \textit{SteamVR Home}~\cite{website:steamvr-home}.
In the figure, some virtual objects in the left eye present completely inverted colors, while others display washed-out hues. Additionally, the Skybox, which projects the 3D panoramic background scene, appears to be absent in the left-eye view. 
We refer to such issues on VR apps' GUI as \textbf{\textit{\issuesFullCap (\issueAbbr)}}.
The inconsistencies not only mislead users with conflicting information but also discourage users from playing the VR application.
Manual playtesting with human testers could be one possible solution to mitigate \issues~\cite{website:meta-vr-playtest-guide}, but it is time-consuming, labor-intensive, and may expose testers to health and safety risks~\cite{paper:vr-sickness-15-mins, paper:sigchi-2020-vr-sickness-symptoms-2}, prompting researchers to develop automated testing methods.

To better understand the \issues in real-world VR apps, 
we first collect \numberOfIssues bug reports of \issues from \numberOfPlatforms VR-related platforms and conduct an empirical analysis to manually analyze their manifestations.
Our findings reveal that the \issues are diversified in both scale and manifestation,
and are closely tied to the semantics of the VR apps' semantics or logic.
For scale, the symptoms span both view level and object level. View-level inconsistencies are global and encompass view displacement, deformation, and view angle discrepancies. Object-level inconsistencies are local and related to object quantity, rendering effects, position, etc. 
For manifestations, we summarize
15 different categories of them.
Instead of causing application crashes or runtime errors, these issues only affect user experience and thus can hardly be detected with regular test oracles.

Recent research has taken steps to address the above-mentioned shortcomings of conventional test oracles and testing approaches in detecting GUI display issues, using deep learning (DL) techniques in automated GUI testing~\cite{paper:owl-eyes, paper:glib, paper:seenomaly, paper:html5-canvas-test}.
These methods model GUI issue detection as a classification problem, augmenting training data by generating abnormal screen captures through modifications of regular GUI screenshots or bug injections into the code. 
Such supervised approaches enable DL classifiers to detect faults effectively and enhance GUI testing efficiency. 
Despite the advancements, they still face the following three challenges in identifying \issues:
(1) \textit{Lack of labeled training data.} It is hard to collect sufficient labeled data for model training, especially data with confirmed \issues. Manual inspection of large amounts of data which might cause cybersickness is inapplicable. Data augmentation methods cannot work well, either, due to the following two challenges.
(2) \textit{Semantic-related manifestations can hardly be captured by pre-defined patterns.} \issues are closely linked to application-specific semantics, making it challenging to be captured using existing pattern-based detection techniques. Besides, current approaches are limited to detecting predefined patterns and can hardly handle unreported symptoms.
(3) \textit{Closed-source VR apps provide limited accessible information for issue detection.} Some approaches supplement issue detectors with internal application data like code and scene configurations~\cite{paper:glib, paper:html5-canvas-test}, while commercial VR apps only expose externally rendered states for \issueSingular detection. 

To address the challenges, in this paper, we propose \textbf{\tool}, an automated testing framework to \toolFullUnderlined in VR apps. \tool relies solely on external rendered states of VR apps and does not require additional information such as code configurations.
Instead of pre-defining detection patterns, \tool reformulates the \issues identification problem into an anomaly detection problem. \tool generates a synthetic right-eye image based on the actual left-eye image and computes distances between the synthetic and the actual right-eye images to detect anomaly\footnote{The model can be used to generate from left to right and vice versa. We take the left-right direction as an example in the rest part of the paper.}.
For the generation of synthetic right-eye images, we propose a \aemodelFullUnderlined (\textbf{\aemodelShort} in short). \aemodelShort captures the complicated but predictable mappings between left-eye and right-eye images. To deal with the spatial shift between left-eye and right-eye images due to object depths in the scene, \aemodelShort integrates monocular depth maps for the left-eye image and the right-eye image respectively as additional inputs. This depth-aware manner empowers \aemodelShort with the crucial spatial context it requires to generate accurate right-eye images.

In summary, we make the following contributions:
\begin{itemize}[leftmargin=*, topsep=2pt, itemsep=2pt]
\item To the best of our knowledge, our work is the first to systematically analyze and detect the \issuesFull (\issues) in real-world VR apps. We build a dataset of bug reports and screenshots with \issues.
\item We construct a large-scale dataset of over 171K VR stereo image screenshots via execution of 288 real-world VR apps.
\item We propose \tool, an automatic testing framework to detect \issues. \tool is empowered by a novel \aemodelFull.
Extensive evaluations show that \tool can effectively detect \issues in real-world VR apps. %
	
\item To facilitate follow-up studies, we release our datasets and \tool, at \\ \href{https://sites.google.com/view/stereoid}{\color{blue}{https://sites.google.com/view/stereoid}}.
\end{itemize}

\section{Empirical Analysis of \issuesCap}
\label{sec:mc-study}

\subsection{Data Collection}
We collect real-world \issues by searching the keywords in reports of VR users or developers from 15 platforms: VR-related online forums and app stores. The 15 platforms include VR online forums (Unity-related forums~\cite{website:unity-discussions, website:unity-forum, website:unity-issue-tracker}, Unreal Engine related forums~\cite{website:ue-forum, website:ue-issue-tracker}), VR app stores and app store forums (Meta Quest App Store~\cite{website:oculus-app-store}, Meta Quest App Lab~\cite{website:oculus-app-lab}, VIVEPORT~\cite{website:viveport}, SideQuest~\cite{website:sidequest}, Steam~\cite{website:steam-app-store-vr}, Meta Community Forums~\cite{website:meta-community-forum}, VIVE Forum~\cite{website:vive-forum}, Steam Community~\cite{website:steam-community}), GitHub~\cite{website:github}, and Stack Overflow~\cite{website:stack-overflow}. To include as many related posts as possible, we start by sampling some posts and analyzing the related keywords, such as \textit{eye render, rendering left eye, rendering right eye, two eyes, both eyes, eyes render difference, inconsistent render (display)}, in the sampled posts. The identified keywords are then used to search for more related posts for keyword analysis. This process iterates until no more new posts and keywords can be found. After the keyword search, we get 3,266 candidate bug reports. As keyword search cannot guarantee that the candidate bug reports are related to \issues, two authors further check the contents of candidate bug reports. 282 distinct bug reports are agreed to be related to \issues by both authors and we finally collect a screenshot dataset of \numberOfIssuesWithScreenshots image pairs with real-world \issues guided by the 282 bug reports.

\subsection{Categorizing the Manifestation of \issuesCap}
\label{sec:category}

\begin{figure*}[h!]
\vspace{-1em}
	\centering  
	\subfigure[Monocular Blindness]{
		\label{fig:view-level-Monocular-Blindness}
		\includegraphics[width=0.18\linewidth]{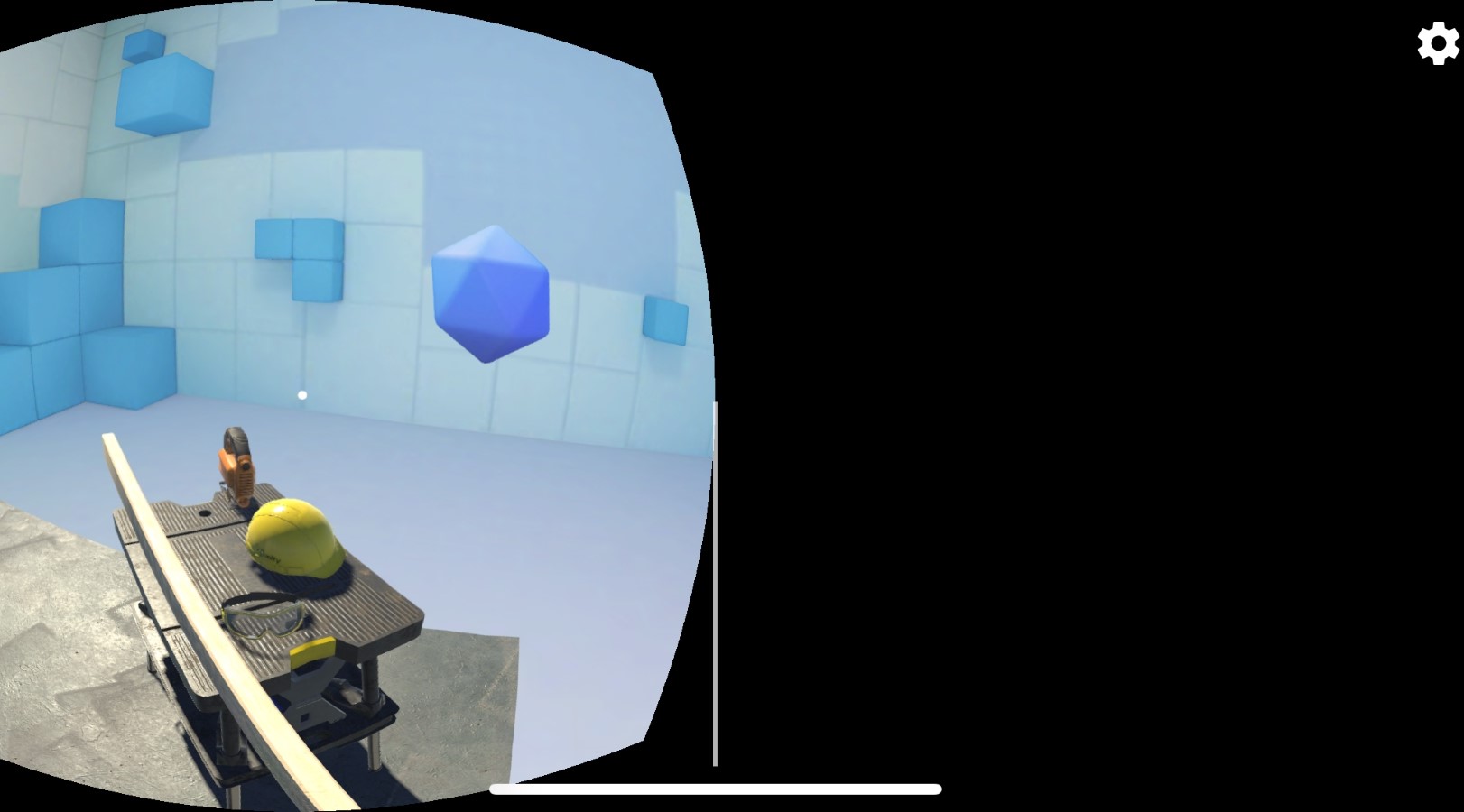}}
	\subfigure[View Misalignment]{
		\label{fig:view-level-View-Misalignment}
		\includegraphics[width=0.18\linewidth]{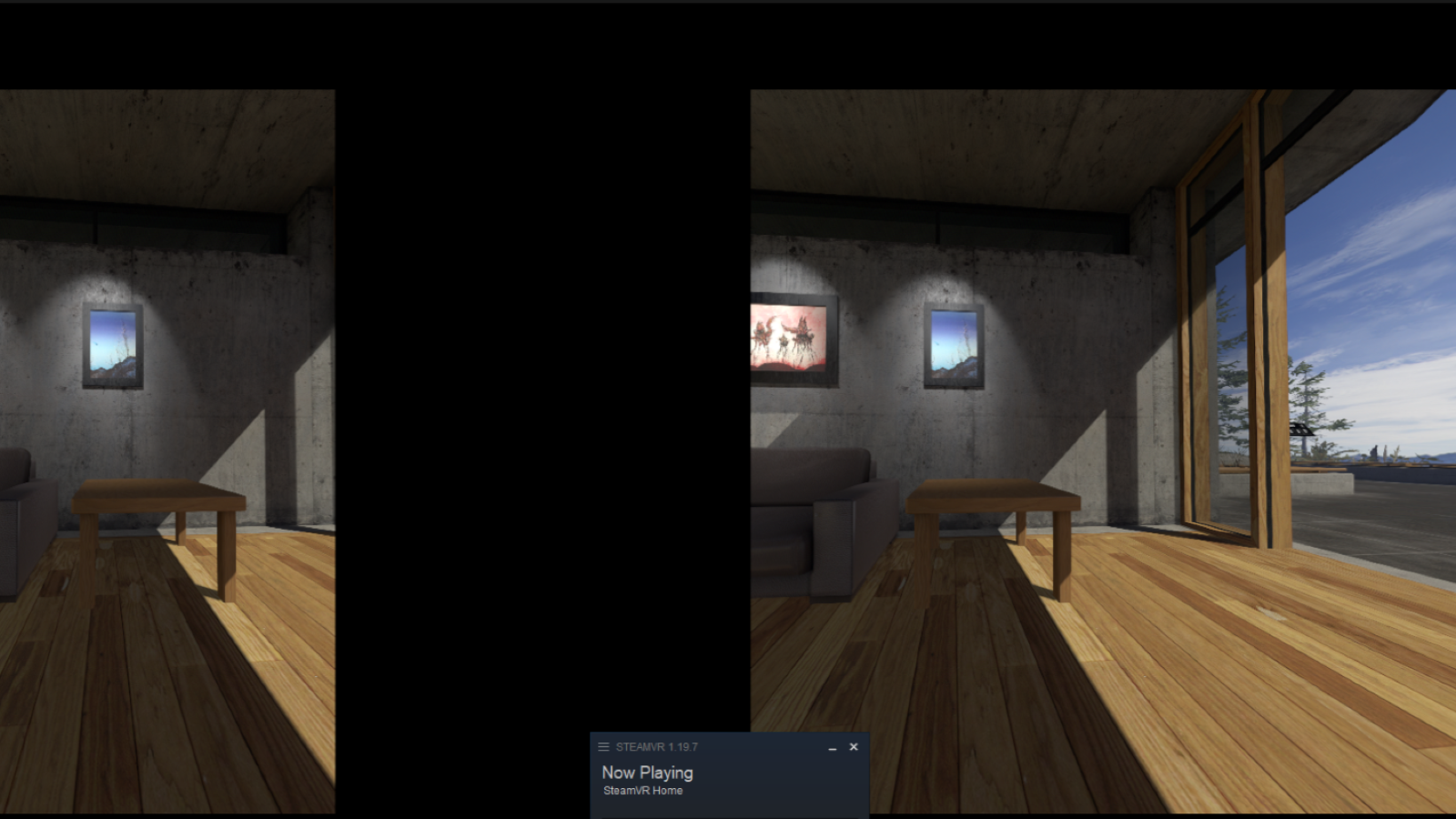}}
	\subfigure[Warped Views]{
		\label{fig:view-level-Wraped-Views}
		\includegraphics[width=0.18\linewidth]{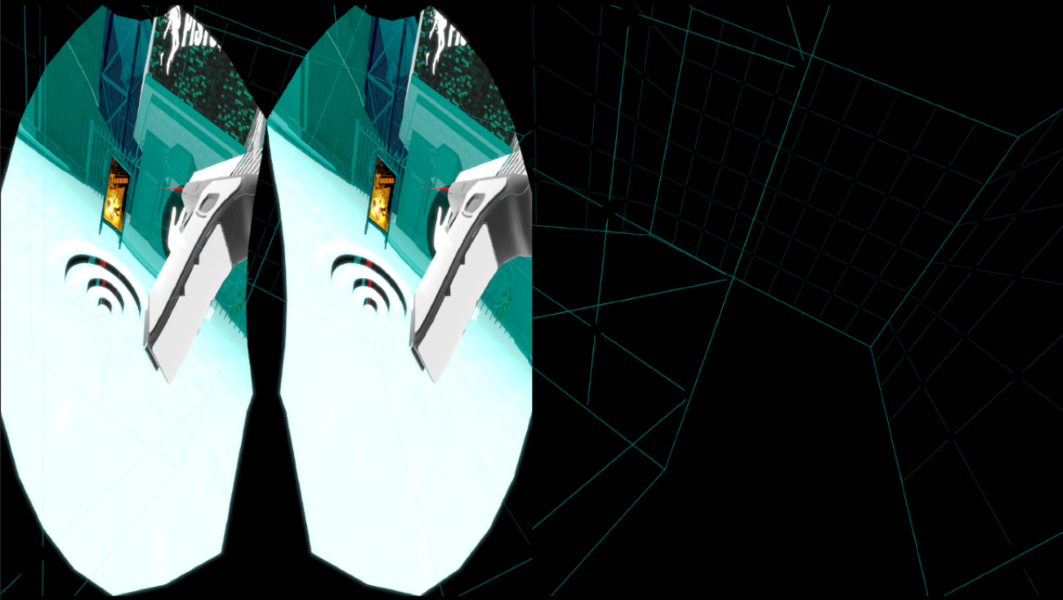}}
  	\subfigure[REI: Lighting and Shadow Discrepancies]{
		\label{fig:Object-level-Lighting-and-Shadow-Discrepancies}
		\includegraphics[width=0.18\linewidth]{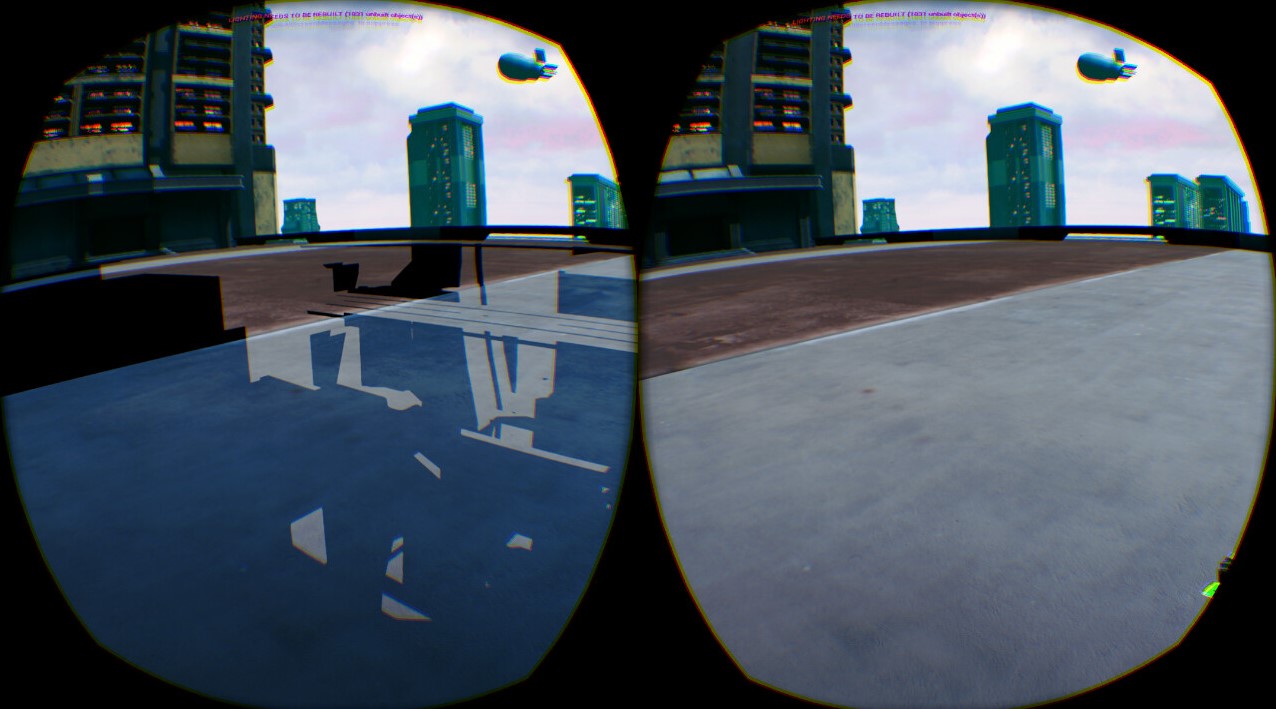}}
	\subfigure[REI: Shader Absence]{
		\label{fig:Object-level-Shader-Absence}
		\includegraphics[width=0.18\linewidth]{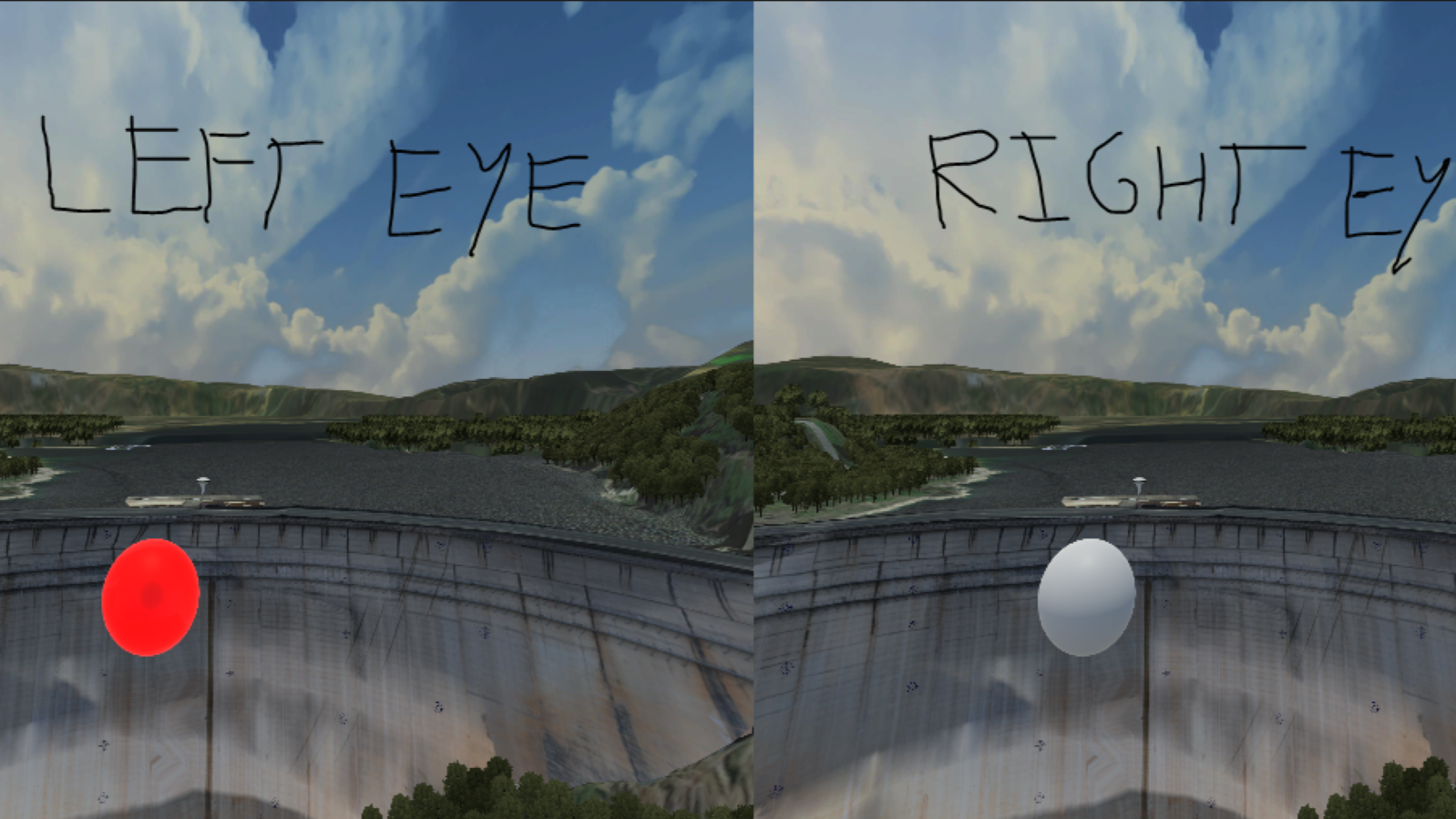}}

	\subfigure[REI: Material or Texture Mismatch]{
		\label{fig:Object-level-Material-or-Texture-Dismatch}
		\includegraphics[width=0.18\linewidth]{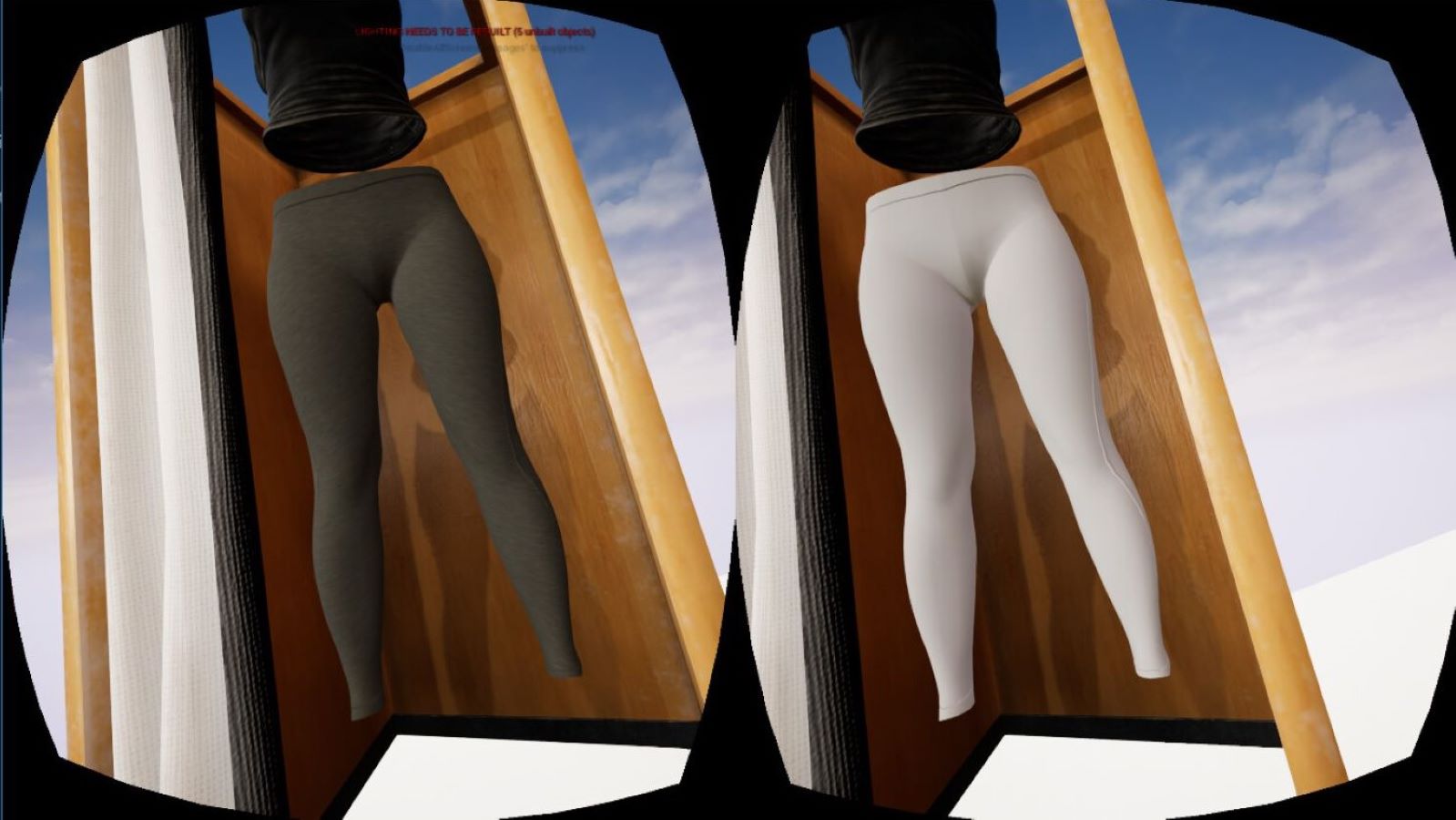}}
	\subfigure[REI: Postprocessing Inconsistency]{
		\label{fig:Object-level-Post-Processing-Inconsistency}
		\includegraphics[width=0.18\linewidth]{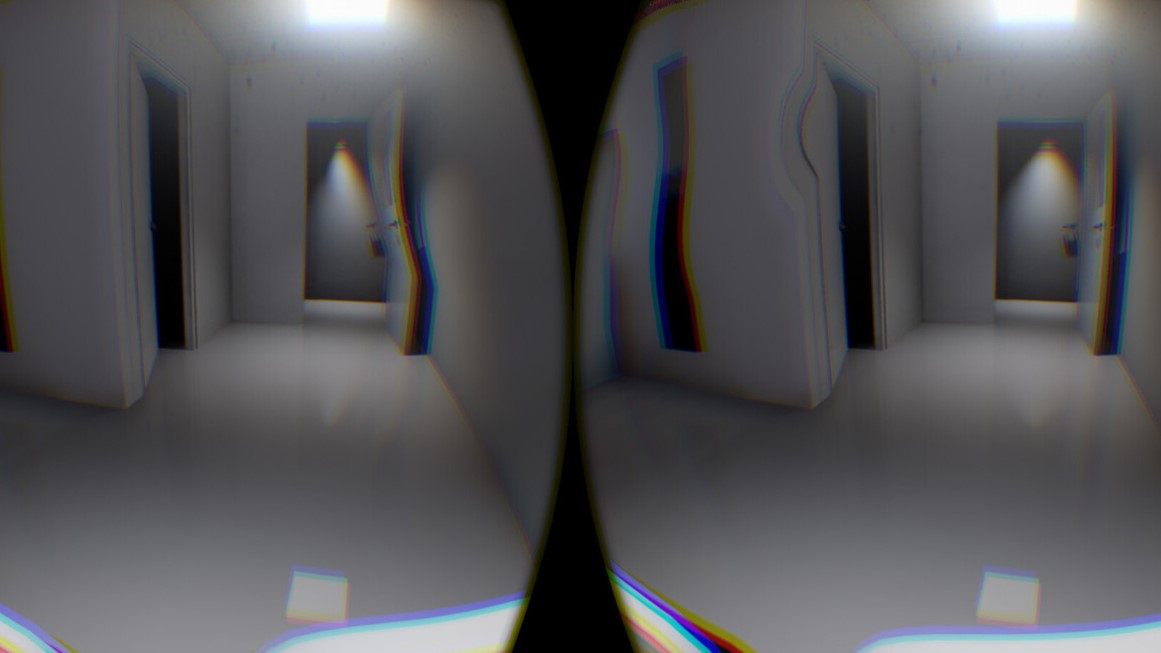}}
  	\subfigure[REI: Particle and Visual Effect Variations]{
		\label{fig:Object-level-Particle-and-Visual-Effect-Variations}
		\includegraphics[width=0.18\linewidth]{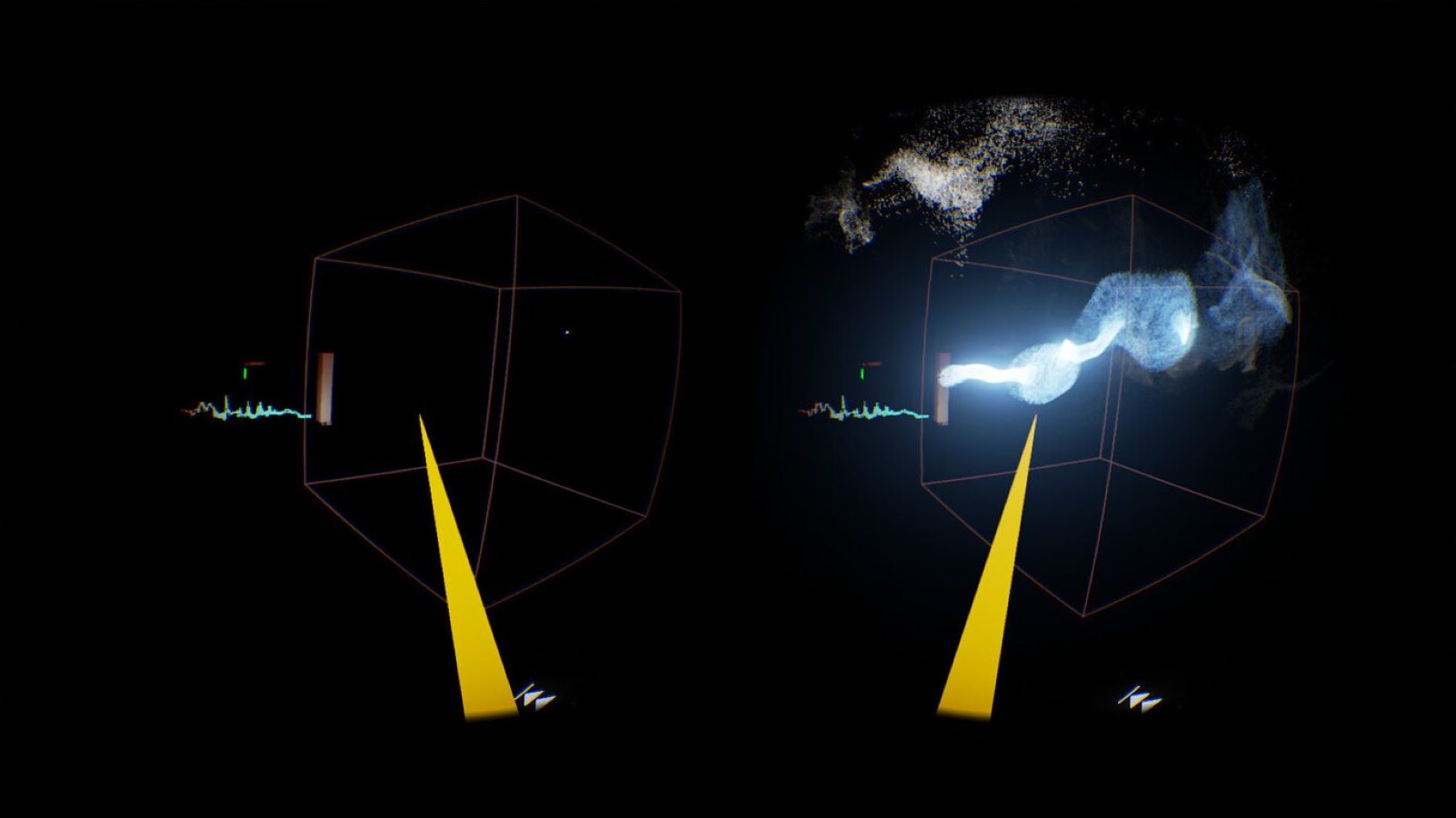}}
  	\subfigure[Object Omission]{
		\label{fig:Object-level-Object-Omission}
		\includegraphics[width=0.18\linewidth]{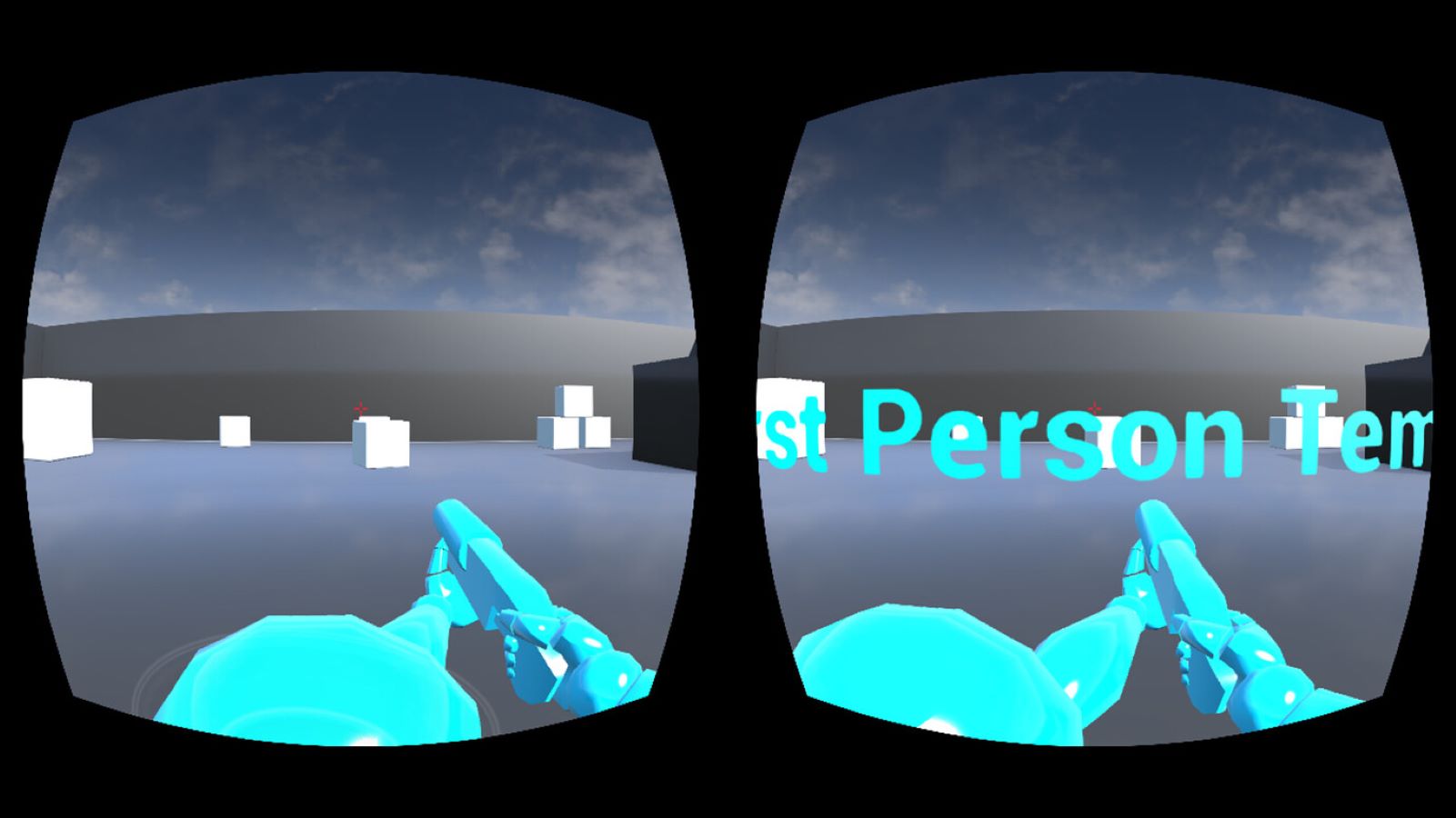}}
	\subfigure[Unilateral Object Rendering]{
		\label{fig:Object-level-Unilateral-Object-Rendering}
		\includegraphics[width=0.18\linewidth]{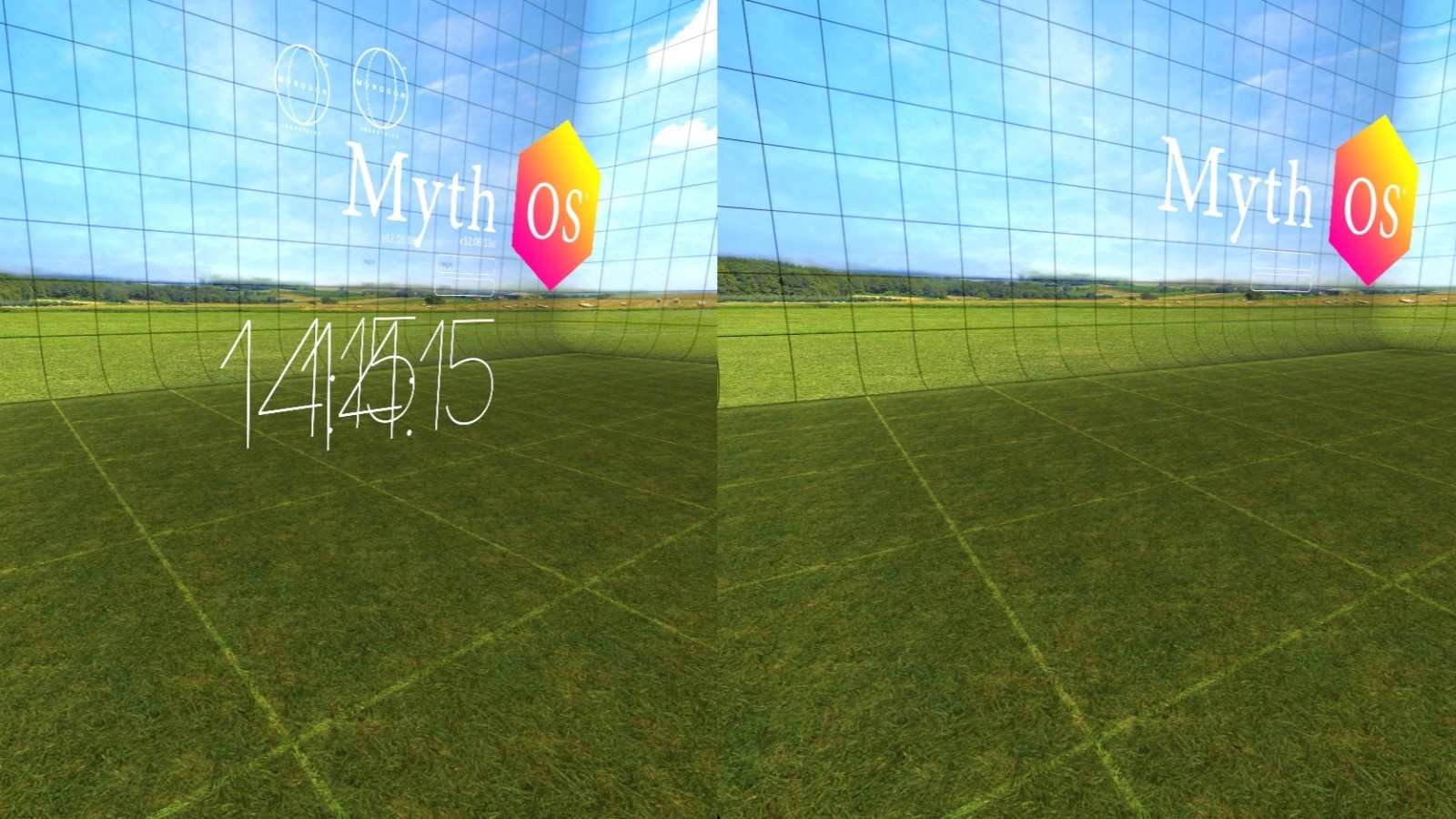}}

	\subfigure[Object Position Discrepancy]{
		\label{fig:Object-level-Object-Position-Discrepancy}
		\includegraphics[width=0.18\linewidth]{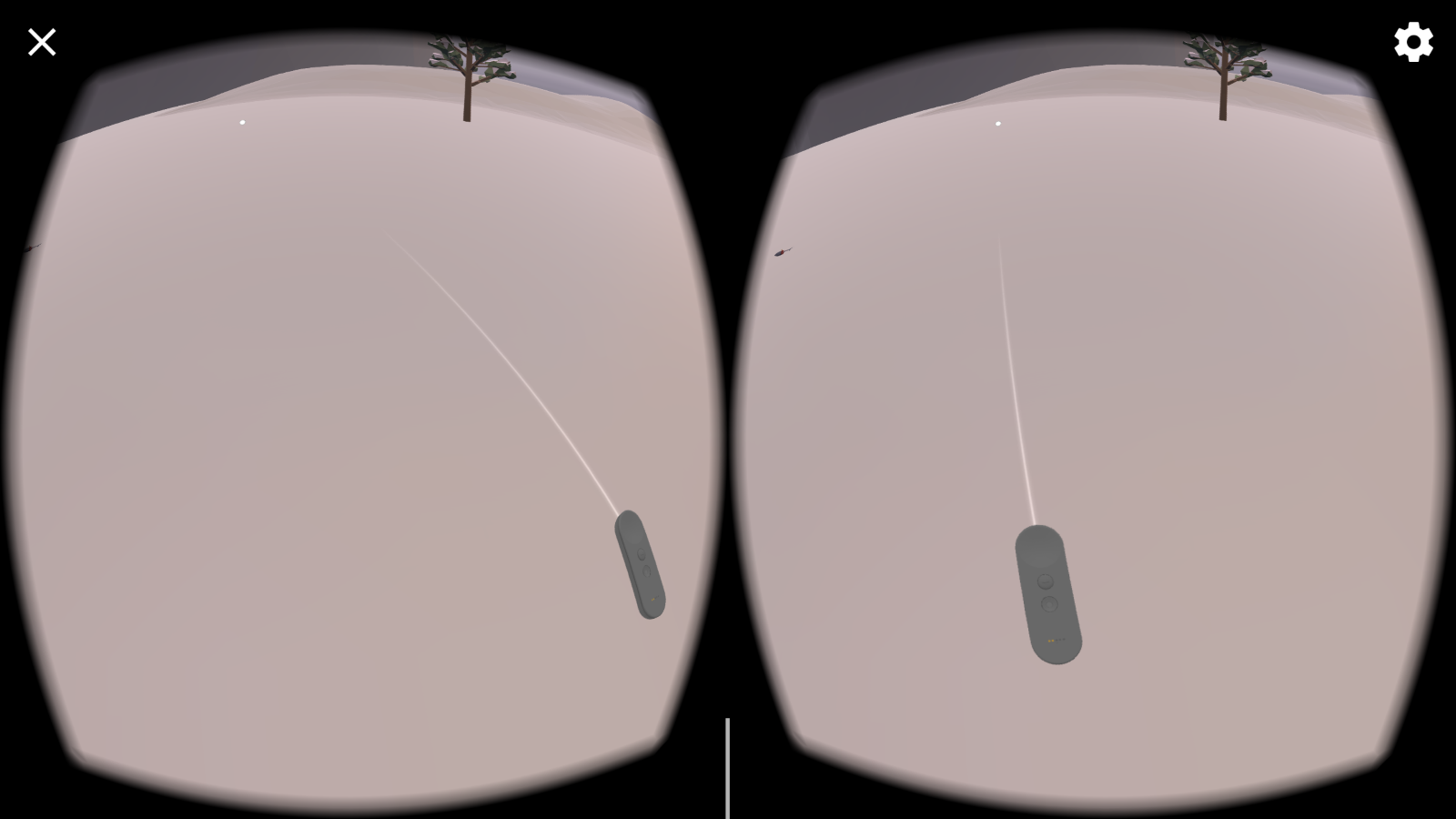}}
  	\subfigure[Object Warping]{
		\label{fig:Object-level-Object-Wraped}
		\includegraphics[width=0.18\linewidth]{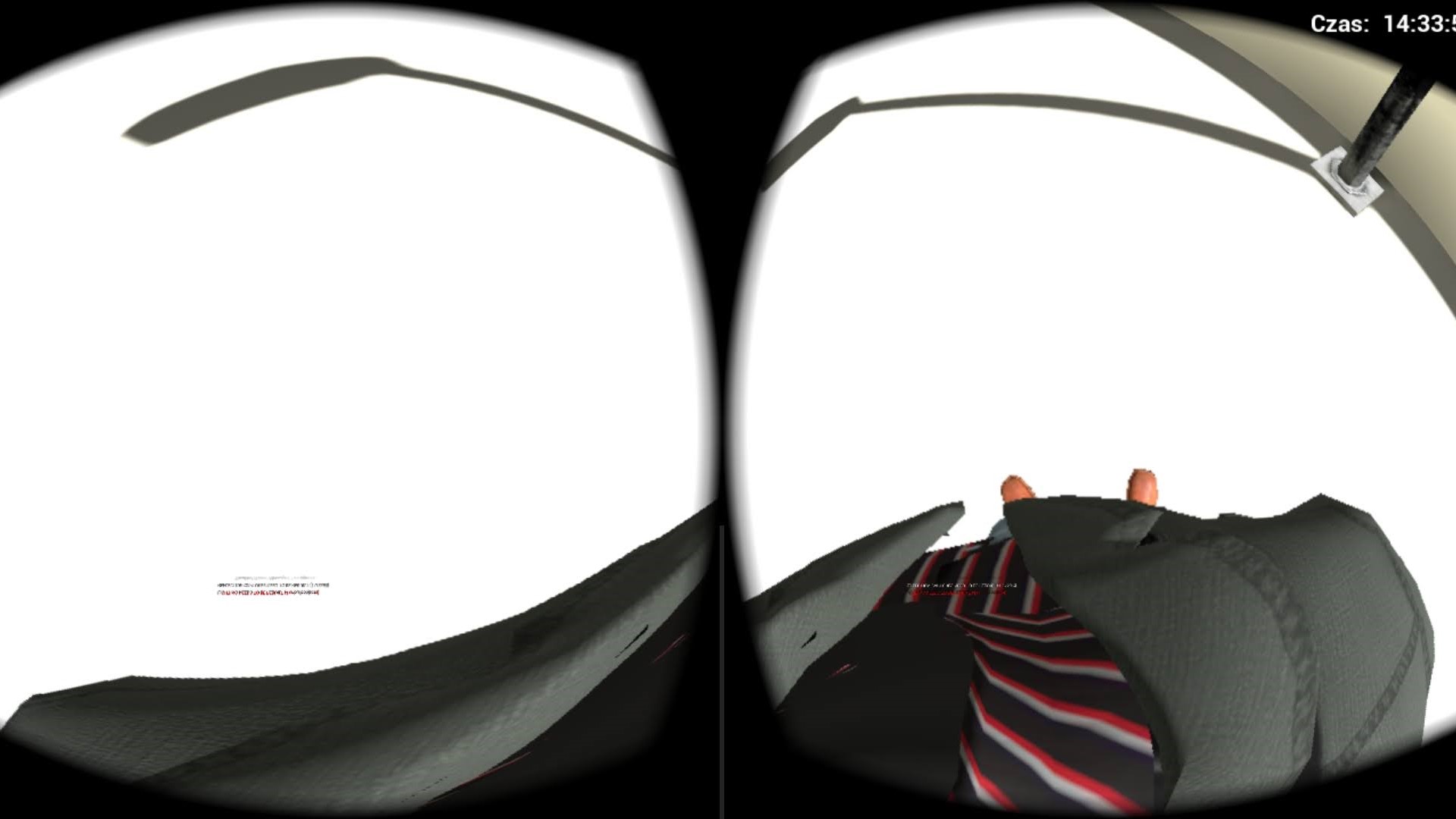}}
  	\subfigure[Level of Detail Inconsistency]{
		\label{fig:Object-level-Level-of-Detail-Inconsistency}
		\includegraphics[width=0.18\linewidth]{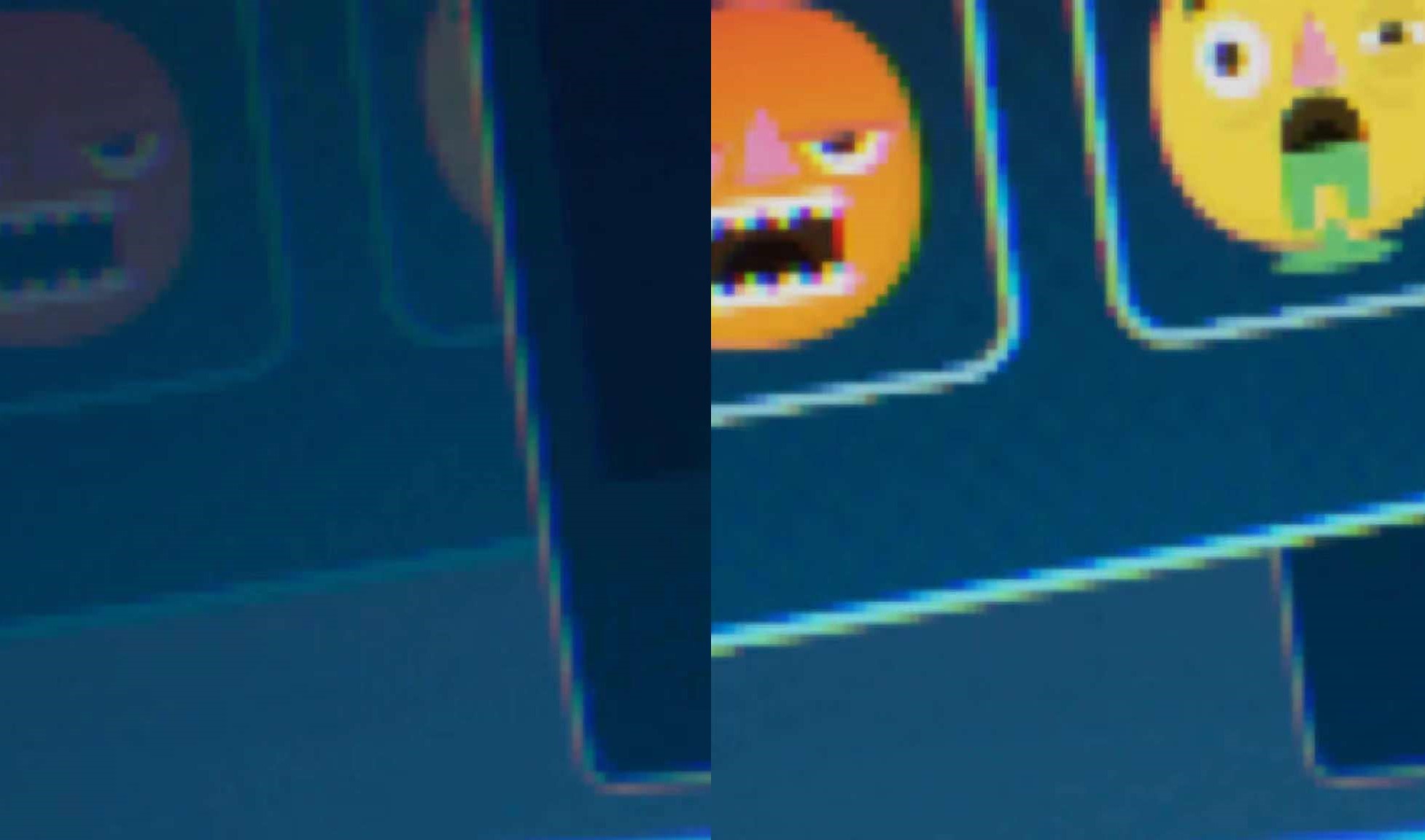}}
  	\subfigure[Partial Object Rendering]{
		\label{fig:Object-level-Partial-Object-Rendering}
		\includegraphics[width=0.18\linewidth]{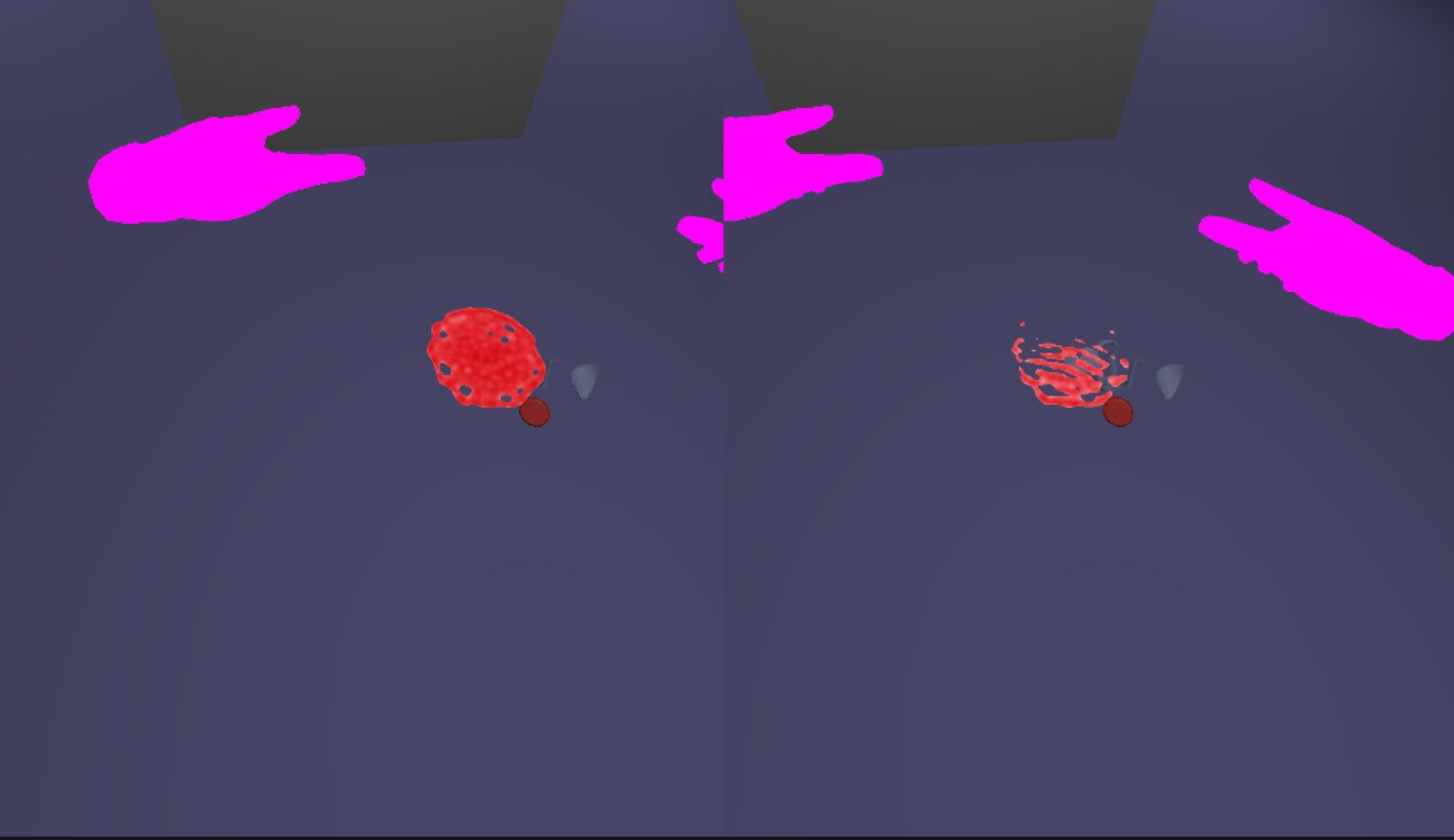}}
	
 \vspace{-1em}
 \caption{Examples of \issuesFull}
	\label{figs:xr-device}
 \vspace{-1em}
\end{figure*}

Following the widely-adopted open coding procedure~\cite{book:open-coding-16}, two authors of this paper, with over three years of software development experience and sufficient domain knowledge on VR, perform further analysis on the manifestation of \issues.
They first individually examine the title, description, discussions, and all uploaded attachments, such as screenshots, of the collected bug reports to understand the specific manifestation of each \issueSingular, and then discuss their results to add/delete, update, or merge codes together. In case of disagreement in the discussion, another three authors are involved in making final decisions. Finally, we reach a consensus on the categorization of \issues and report our findings as follows\footnote{The raw data can be found in our dataset.}.
The SVI issues are classified
into two types: view-level (global) and object-level (local) inconsistencies. View-level inconsistencies refer to issues that impact the entire view, while object-level inconsistencies pertain to inaccurate rendering of single objects. Note that some bug reports have multiple manifestations and thus they belong to multiple categories. We classify bug reports with vague descriptions and manifestations with low occurrences into the ``Other'' category.

We present the refined symptom categories of \issues in the rest of the section.

\subsubsection{View-Level Inconsistencies}
~

\textbf{Monocular Blindness (17\%)}. This category covers issues where one eye fails to render any visual information, resulting in a blank or black screen for that eye. Figure~\ref{fig:view-level-Monocular-Blindness} shows that the right-eye image is not rendered.

\textbf{View Misalignment (6\%)}. This category pertains to issues where the left-eye and right-eye views are misaligned, causing a noticeable offset between the two views. Figure~\ref{fig:view-level-View-Misalignment} depicts the left-eye view as distinctly offset to the left, obscuring its full view for the user.

\textbf{Warped Views (3\%)}. This category includes issues where the visual information displayed to each eye is distorted or stretched, resulting in a warped or unnatural view. Unlike view misalignment, warped views can also cause object distortion and alter the overall proportions of the view. Figure~\ref{fig:view-level-Wraped-Views} shows the left-eye and right-eye images rendered normally but with abnormal compression and ratio. Ultimately, only the left-eye view is displayed, leaving the right-eye view dark.

\textbf{Asymmetric Viewing Angles (1\%)}. This category involves issues where the viewing angles of the left and right eye views are inconsistent, or one eye's view is flipped or skewed. For example, one bug report ~\cite{website:vive-eyes-displacements} shows the following manifestations: ``\textit{when each eye gazes in a different direction, the left eye sees the object from the side while the right eye sees it from the front.}'' This difference in perspective, despite their similar positions in the field of view, leads to double vision and dizziness.

\subsubsection{Object-Level Inconsistencies}
~

\textbf{Rendering Effect Inconsistencies (REI) (29\%)}. This category covers various types of inconsistencies related to the rendering of graphical effects as follows.
\begin{itemize}[leftmargin=*, topsep=2pt, itemsep=2pt]
	\item \textit{Lighting and Shadow Discrepancies (11\%)}: differences or absences in lighting, shadows, or reflections between the eye views. In Figure~\ref{fig:Object-level-Lighting-and-Shadow-Discrepancies}, there is an inconsistency in the shading between the left and right eyes.
 	\item \textit{Shader Absence (6\%)}: missing or incomplete visual effects in one eye view due to shaders not rendering. In Figure~\ref{fig:Object-level-Shader-Absence}, a user-defined shader for a red highlight material is rendered solely in the left eye, while being absent in the right eye.
  	\item \textit{Material or Texture Mismatch (6\%)}: missing or mismatched textures between the two eyes. In Figure~\ref{fig:Object-level-Material-or-Texture-Dismatch}, the materials of clothing for the left and right eyes differ, with the left eye featuring black pants and the right eye featuring
   white ones.
   	\item \textit{Post-Processing Inconsistency (3\%)}: inconsistent post-processing effects between the two eyes. Post-processing effects offer several specific rendering effects with little latency for developers~\cite{website:unity-manual-post-processing, website:unity-post-processing-list}. Inconsistent post-processing effects can lead to vignettes, blurring, or ghosting in the monocular view.
    In Figure~\ref{fig:Object-level-Post-Processing-Inconsistency}, there is a disturbance and blurriness in the center of the left and right eye views, as they appear at different positions. 
 	\item \textit{Particle and Visual Effect Variations (3\%)}.
    The Visual Effect Graph~\cite{website:unity-veg} creates a particle system to simulate particle behavior, generating Visual Effects like varying appearances, explosions, or smoke within a single view, greatly enhancing immersion and gameplay. Particle effects can be considered specialized effects generated by a particle system. Inconsistency of such effects can cause variations in phenomena such as smoke, sparks, shooting stars, clouds, dust, and more in both eyes.
    For instance, in Figure~\ref{fig:Object-level-Particle-and-Visual-Effect-Variations}, GPU particles in the view are only rendered on the right eye, leaving the corresponding part of the left eye black.
\end{itemize}

\textbf{Object Omission (20\%).} This category pertains to issues where some objects are not rendered in either eye view, leading to a loss of important visual information. E.g., in Figure~\ref{fig:Object-level-Object-Omission}, the blue text is only visible in the right eye, indicating the omission of the text object in the left eye view.

\textbf{Unilateral Object Rendering (7\%)}. This category includes repeated objects or foreign elements that shouldn't exist in the monocular view, leading to an uneven visual experience for the user. For example, the time text shown in Figure~\ref{fig:Object-level-Unilateral-Object-Rendering} is repeated in the left eye.

\textbf{Object Position Discrepancy (7\%)}. This category pertains to issues where the position or orientation of objects differs between the left and right eye views. For instance, in Figure~\ref{fig:Object-level-Object-Position-Discrepancy} the controller appears in a different position in each eye.

\textbf{Object Warping (2\%)}. This category encompasses issues where some objects appear distorted or stretched in one eye view. In Figure~\ref{fig:Object-level-Object-Wraped}, the shadow looks bigger in the left image and the jacket is stretched.

\textbf{Level of Detail (LOD) Inconsistency (2\%)}. 
The level of detail refers to the complexity of 3D VR models which can be decreased for distant or dynamically changing objects.
This category includes issues where the level of detail differs between the two eyes' views, such as level of detail inconsistencies or anti-aliasing problems.
In one bug report from the Unity Forum~\cite{website:eye-different-lod}, the VR application displays different levels of detail for each eye, leading to disparities at specific distances. For example, in Figure~\ref{fig:Object-level-Level-of-Detail-Inconsistency}, when two planes intersect in front of the user, it creates an erroneous perception of varying distances between the planes and the L/R  perspectives, resulting in incorrect depth perception for users.

\textbf{Partial Object Rendering (1\%)}. This category involves issues where some objects are only partially rendered.
Take Figure~\ref{fig:Object-level-Partial-Object-Rendering} as an example, the red fluid is displayed normally in one eye, but becomes mutilated and incomplete in the other eye~\cite{website:right-eye-discrepancies}.

\subsection{Challenges to Automatic Detection of \issuesCap}
\label{subsec:empirical-ad-challenges}

\issues adversely affect user experience without causing application crashes or throwing errors, making the detection of them difficult with regular test oracles.
There exist prior studies~\cite{paper:owl-eyes, paper:glib, paper:seenomaly, paper:html5-canvas-test} on automated GUI test oracles that employ DL techniques for identifying abnormal GUI states in mobile, web apps, and games. 
These methods typically formulate the process of GUI issue detection as a classification task. To enhance the training data, they generate anomalous screen captures by modifying standard GUI screenshots or by injecting bugs directly into the codebase.
Further details are presented in Section~\ref{sec:related-work}.
However, despite these progressive advancements, the
existing methods struggle to accurately identify \issues and face the following challenges. 

\textbf{Challenge 1: Lack of training data.} While the concerns of \issues widely exist in the forums of VR apps, few bug reports from users include the described problematic images. This is evidenced by only 108 (<40\%) image pairs extracted from the 282 bug reports. For data augmentation methods to enrich the dataset, both image-based and code-based methods rely on manually identified glitch or bug patterns, which exhibit similar manifestations across different apps, such as random noise, overexposure, and black borders. However, as introduced in Section~\ref{sec:category}, the symptoms of \issues are strongly related to application semantics and exhibit diverse manifestations across different scenarios, making it challenging to capture them with a set of general image patterns.

\textbf{Challenge 2: Semantic-related manifestations can hardly be captured by pre-defined patterns.}
To augment with, neither image-based nor code-based data augmentation can work well even on one specific image pattern.
For image-based methods, complex S3D rendering for VR is hard to manipulate and mimic feasibly according to a predefined image pattern as real-world \issues, while maintaining the consistency of the rest part.
We observe that the root cause of \issues is more complex than several lines of code changes, and the bugs arise from multiple sources spanning apps, developing engines, runtime, and related libraries.
Thus, it is also hard to inject bugs and conduct code-based data augmentation.

\textbf{Challenge 3: Closed-source VR apps provide limited accessible information for issue detection.} Previous approaches require semantics information derived from internal application data such as code, assets, and scene configurations. The internal application data, however, can hardly be accessed for commercial VR apps. Limited input information obtained from commercial VR apps hinders previous approaches to effectively detect \issues in practice.

\begin{figure*}[t!] 
	\centering 
	\includegraphics[width=\textwidth]{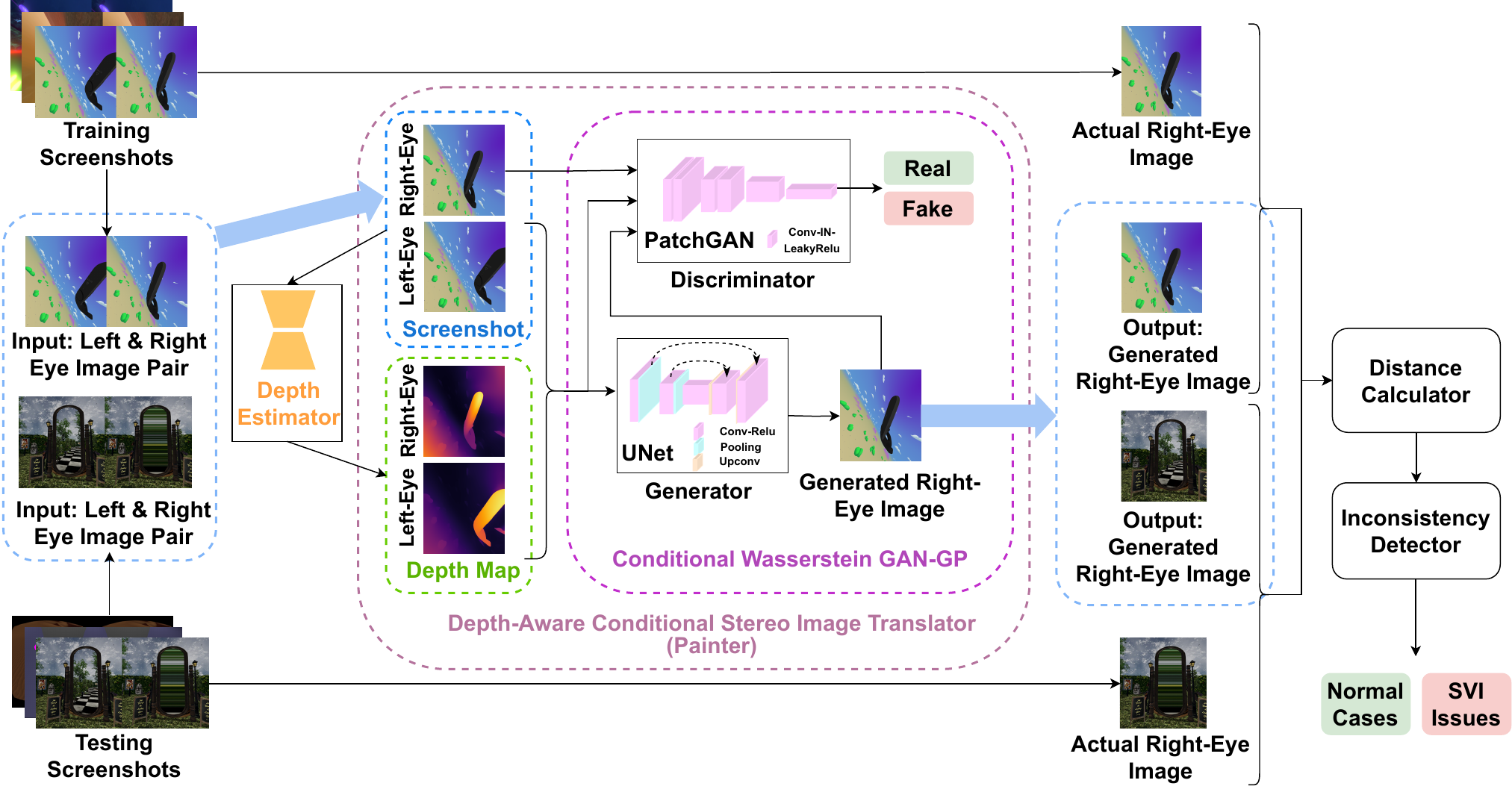} 
  \vspace{-2em}
	\caption{\textblue{Overview of \tool}}
	\label{fig:stereoid}
        \vspace{-1.5em}
\end{figure*}

\section{Approach: \tool}
\label{sec:approach}
In this section, we elaborate on the proposed
automatic testing framework, named \tool, for identifying  \issues.
As illustrated in Fig.~\ref{fig:stereoid}, \tool mainly includes four components, including \textit{monocular depth estimator}, \textit{depth-aware left-right-eye image translator}, \textit{distance calculator}, and \textit{inconsistency detector}.
Given a dataset of paired stereo screenshots with left-eye and right-eye images, the \textit{\aemodelFull} tries to learn the mapping relations between the rendered GUI screenshots of the two eyes to generate
synthetic right-eye images given the
left-eye image. 
To make the generation a depth-aware process 
to calculate and involve the 2D representations (referred to as depth maps) that encode the distance between the viewer and virtual objects in the scene for each pixel for both eyes.
A distance metric is then computed by the \textit{distance calculator} between the generated synthetic right-eye image and the real right-eye image, which serves as the basis for the inconsistency detection procedure.
Outliers in this metric space are finally identified by \textit{inconsistency detector} as
\issues.
The \tool framework is unsupervised and does not need any annotated \issues. 

\begin{figure}[t!]
	\centering  
	\subfigure[Original stereoscopic screenshot (left: left-eye view, right: right-eye view)]{
		\label{subfig:left-right-mapping}
		\includegraphics[width=0.33\linewidth]{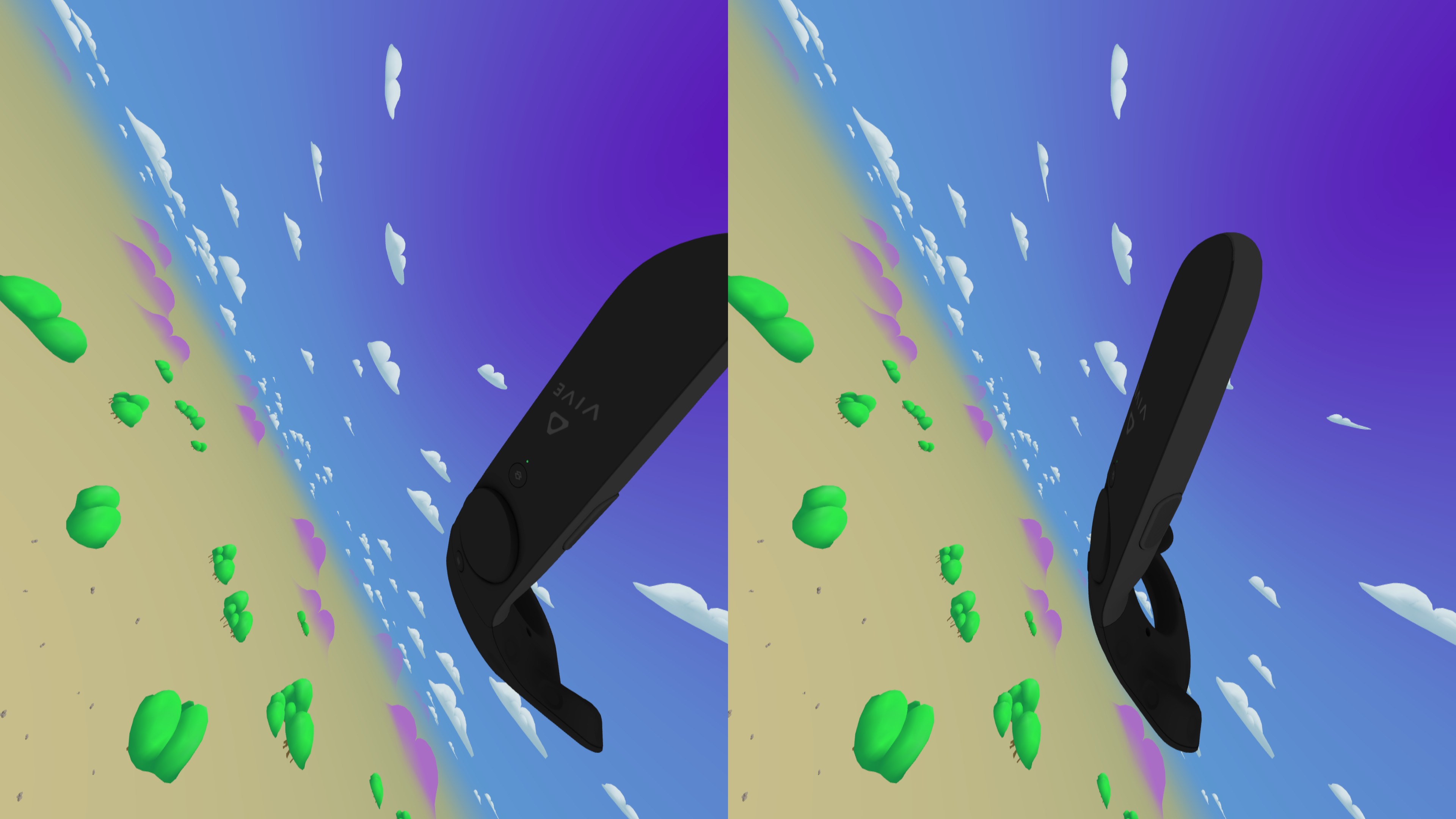}}
	\hspace{5mm}
	\subfigure[Depth map generated by monocular depth estimator (Section~\ref{subsec:depth-estimator})]{
		\label{subfig:depth-map}
		\includegraphics[width=0.33\linewidth]{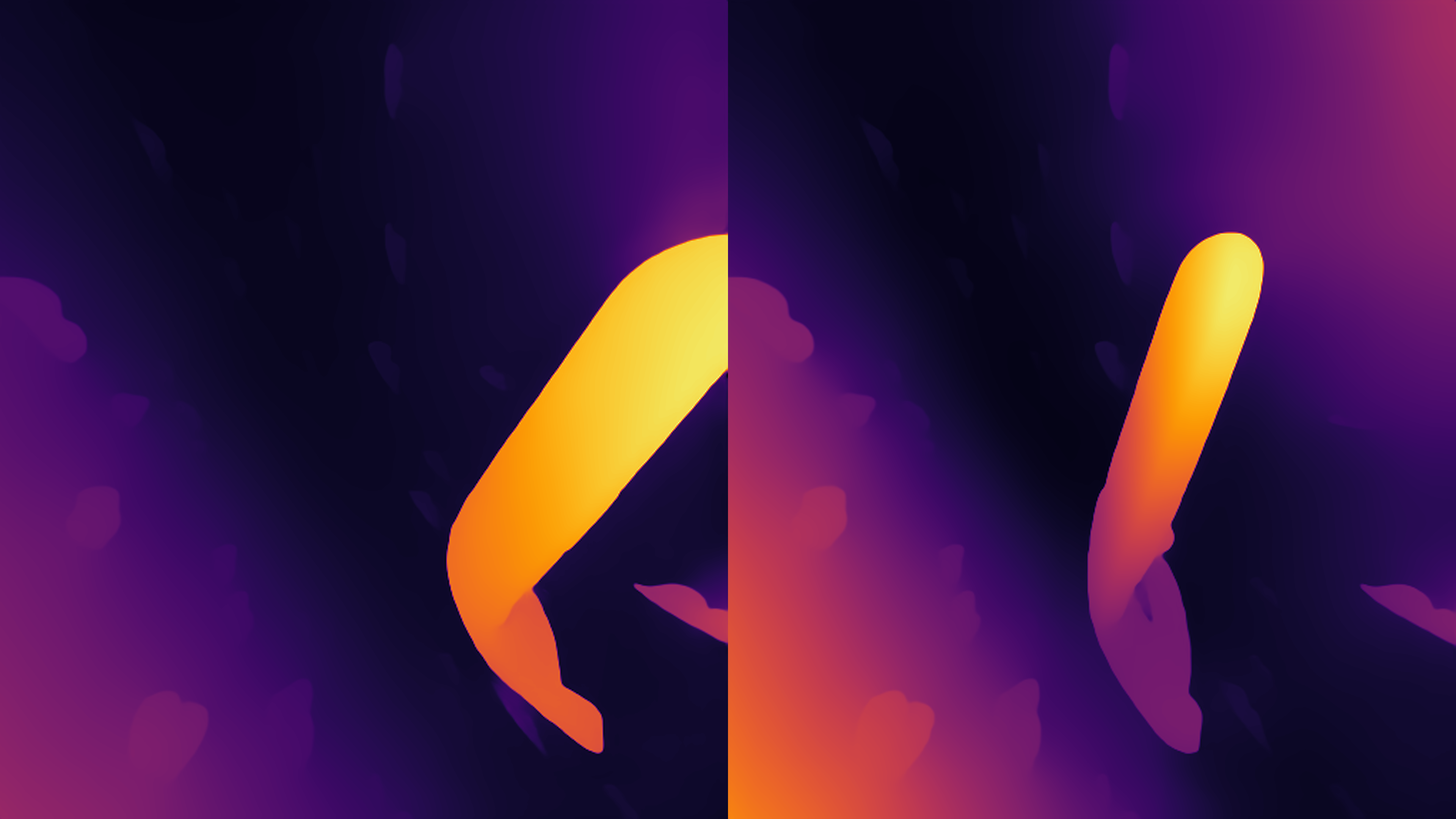}}
  	\vspace{-1em}
        \caption{Illustration of the stereoscopic screenshot and the corresponding depth map\protect\footnotemark}
	\label{figs:left-right-depth-map}
 \vspace{-1em}
\end{figure}

\footnotetext{The depth map assigns each pixel color based on the actual distance from the viewer (camera) to the object, with surfaces closer to the viewpoint appearing darker in color, and those farther away appearing brighter~\cite{paper:conf/cvpr/JeonPCPBTK15}.}

\subsection{Monocular Depth Estimator}
\label{subsec:depth-estimator}

In \tool, we consider the depth information inherent
in the stereo-mapping generation process in S3D VR scenarios. As shown in Fig.~\ref{subfig:left-right-mapping}, the relationship between the left-eye and right-eye images essentially relates to the shifts of
virtual objects on the scene, and the shift
is strongly related to the \textit{depth} of each object (i.e., the distance between the object and the camera/viewer). 
To accurately synthesize the right-eye image given the left-eye image, we propose to involve
depth maps as spatial context input. Specifically, we design the monocular depth estimator component for computing the depth maps of left-eye and right-eye images, respectively, before image generation.
The proposed component incorporates $BEiT_{512}-L$~\cite{ranftl2020towards}, a robust monocular relative depth estimation model. 
$BEiT_{512}-L$ employs a Transformer-based architecture which enables it to learn and extract deep features from input images. 
The model has been trained with multi-objective optimization for image classification on up to 12 datasets. 
This allows the accurate generation of depth maps, even when the depth varies across a wide range in a single scene. Fig.~\ref{figs:left-right-depth-map}(b) illustrates
the depth maps generated by our monocular depth estimator for Figure~\ref{subfig:left-right-mapping}, in which each depth map is generated from one monocular image.

\subsection{\aemodelFullCap}
\label{sec:painter}
\subsubsection{Overview of \aemodelShortCap}

With the depth maps generated by the monocular depth estimator, \aemodelFull (\aemodelShortCap in short) is able to learn the complex mappings from the provided context and generate an expected corresponding stereoscopic image from a monocular-eye view.

The principal architecture of \aemodelFull is a conditional variant of Generative Adversarial Networks (GAN)~\cite{mirza2014conditional}, specifically a Conditional Wasserstein GAN~\cite{arjovsky2017wasserstein} with Gradient Penalty~\cite{gulrajani2017improved}, to generate
right-eye images from corresponding left-eye images and their respective depth maps generated by the monocular depth estimator. 

The model may not be stable, i.e., sometimes it generates low-quality samples or fails to converge.  The gradient penalty (GP)~\cite{gulrajani2017improved} is proposed to enhance the stability of training for the Wasserstein GAN framework.
To ensure the stability of the training process, our model employs a gradient penalty that is designed to penalize the norm of the gradient of the output of the discriminator $D$ with respect to its input. This mechanism encourages the gradient norm to be close to 1, effectively enforcing the Lipschitz constraint without the potential drawbacks of weight clipping. This approach ensures a stable learning process and precludes the model from potential divergence, thereby increasing the overall robustness of \tool.

\subsubsection{The Generator Architecture of \aemodelShortCap.}

In our proposed model, the synthesis of the right-eye images is carried out by a generator constructed around a U-Net architecture~\cite{paper:conf/miccai/RonnebergerFB15}). This architecture is a prominent choice in image synthesis due to its capacity to capture intricate details and retain context through its design. The U-Net incorporates an encoder-decoder framework complemented by skip connections between mirrored layers in the encoder and decoder blocks. This unique configuration facilitates the transfer of low-level, localized information directly across the network, thereby enhancing the quality of the synthesized output.

The generator is structured to accept three inputs: a left-eye image and depth maps of both eyes. These inputs are aggregated along the channel dimension, resulting in a nine-channel tensor which constitutes the initial input to the U-Net.

The architectural design of the generator includes numerous layers. Each layer is composed of two $3 \times 3$ convolutions, each succeeded by a Rectified Linear Unit (ReLU) activation function, and a $2 \times 2$ max pooling operation with a stride of 2 for downscaling. This structure persists until the model reaches the lowest resolution point in the encoding path.

Following this, the decoding path, or upscaling process, is initiated. Each step in this phase includes an up-convolution operation, subsequent concatenation with the correspondingly deep feature map from the encoding path, and a pair of $3 \times 3$ convolutions, each succeeded by a ReLU. The upscaling process continues until an output image of equivalent dimensions to the input left-eye image is generated, representing the synthetic right-eye image.

The number of features per layer in the generator is governed by a hyperparameter denoted as \texttt{ngf}, which in our implementation is set to 64. As the network transitions from the encoding to the decoding path, this parameter value doubles and halves respectively at each layer, ensuring a balanced distribution of features across the network.

\subsubsection{The Discriminator Architecture of \aemodelShortCap.}

The underlying discriminator in our conditional Wasserstein generative adversarial network harnesses a Markovian architectural design, which is known as PatchGAN~\cite{isola2017image}. This specialized architecture operates on segments of the image, specifically patches of dimensions $N \times N$, enabling it to independently classify each patch as either real or synthetic. This patch-based approach bolsters the capability of the network to synthesize images exhibiting sharper details and textures, thereby focusing on localized structural nuances as opposed to a global perspective.

Our discriminator is designed to accept a twelve-channel input: a composition of the nine-channel tensor provided to the generator (comprising a left-eye image and two depth maps), in addition to the three-channel output from the generator, representing the synthetic right-eye image. Each discriminator input corresponds to a $70 \times 70$ patch of an image.

The structure of the discriminator is characterized by a series of 
layers, with each layer composed of a Convolution-BatchNorm-LeakyReLU sequence. This number of layers is preset to three in our implementation, and the LeakyReLU activation function employs a slope parameter of 0.2 for negative input values. The number of features per layer
commences at 64 and subsequently doubles after each layer.

In the context of enhancing model learning stability, our design incorporates Wasserstein distance~\cite{arjovsky2017wasserstein} coupled with a gradient penalty~\cite{gulrajani2017improved}. This mechanism ensures unit norm gradients, thereby fostering robust learning dynamics and alleviating prevalent issues such as mode collapse often encountered in conventional GAN designs. The magnitude of this gradient penalty is modulated by a hyperparameter, $\lambda_{gp}$, which is preset to a value of ten, adhering to widely adhered standards.

The final aspect of the discriminator architecture revolves around maintaining equilibrium in the training dynamics between the generator and the discriminator. This balance is governed by the \textit{critic\_iterations} parameter, specifying the ratio of discriminator iterations per generator iteration. Set to a value of 5 in our model, this parameter ensures that the critic is adequately trained prior to each update in the generator, thereby upholding a balanced training regimen.

\subsubsection{Objective Functions and Training Approach}

Our model employs a conditional Wasserstein GAN~\cite{arjovsky2017wasserstein} with gradient penalty~\cite{gulrajani2017improved} for the synthesis of right-eye images conditioned on the left-eye image and respective depth maps.

The training methodology necessitates a cyclic optimization of the generator and discriminator with the aim of minimizing an ensemble of loss functions. These functions include the Wasserstein GAN loss ($L_{WGAN}$), the L1 loss ($L_{1}$), and a weighted Mean Squared Error (MSE) loss ($L_{WMSE}$).

\textbf{Wasserstein GAN Loss with Gradient Penalty.} The $L_{WGAN}$ is designed to quantify the distance between the distributions of authentic and synthesized images. Incorporation of the gradient penalty facilitates the enforcement of Lipschitz continuity, yielding stability to the GAN training procedure. The loss is calculated for both real and generated images.
\begin{equation}
\begin{split}
    L_{WGAN}  & = \mathbb{E}_{x \sim P_{\text{real}}} [D(x)] - \mathbb{E}_{z \sim P_{z}} [D(G(z))] + \lambda E_{\hat{x}}[(||\nabla_{\hat{x}}D(\hat{x})||_2 - 1)^2],
\end{split}
\end{equation}

where $P_{\text{real}}$ and $P_{z}$ represent the data and generator distributions, $D$ is the discriminator, and $G$ is the generator. $\hat{x}$ denotes a randomly sampled point along the straight line between a real and a generated data point, $\lambda$ is a hyperparameter that controls the penalty strength, and $||\cdot||_2$ is the Euclidean norm.

\textbf{L1 Loss.} The ${L1}$ loss encourages the generated right-eye images to approximate the ground truth in the absolute difference sense.
\begin{equation}
L_{1} = \mathbb{E}_{x, y \sim P_{\text{data}}} [|y - G(x)|_1],
\end{equation}

where $x$ and $y$ are the input and corresponding target images.

\textbf{Weighted MSE Loss.} The $L_{WMSE}$ loss computes a weighted pixel-level squared difference between the real and generated right-eye images.
\begin{equation}
L_{WMSE} = \mathbb{E}_{x, y \sim P_{\text{data}}} [w(x, y) \cdot (y - G(x))^2],
\end{equation}

where $w(x, y)$ indicates the weight, which is determined by the absolute difference between the real and generated images.
\begin{equation}
w(x, y) = \frac{1}{1+e^{-|y - G(x)|_1}}.
\end{equation}

\textbf{Optimization Procedure.}

The total loss $L$ for the generator is defined as:
\begin{equation}
L = L_{WGAN} + \alpha L_{1} + \beta L_{WMSE},
\end{equation}

where $\alpha$ and $\beta$ are hyperparameters that modulate the contribution of each loss term. The model employs the Adam optimizer, iterating between updating the generator and the discriminator.
\begin{align}
G^{} = \arg\min_{G} L(G, D), \
D^{} = \arg\max_{D} L(G, D),
\end{align}

where $G^{}$ and $D^{}$ represent the optimal generator and discriminator parameters.

\subsection{Inconsistency Detection}

From the empirical analysis, we find that \issues exhibit a wide array of manifestations, some of which may be unknown or unexpected, making it infeasible to learn the ``wrong'' patterns. Additionally, \issues are naturally rare compared to normal behavior, leading to a scarcity of positive samples for supervised training. 
However, \issues are expected to inhabit less dense regions of the feature space and have distinctive attribute values.
Therefore, we formalize the whole testing problem as an anomaly detection process, which focuses on learning the ``right'' patterns and flagging deviations from these patterns as potential issues.

\subsubsection{Distance Calculator}
The objective of our inconsistency detector is to identify the most discrepant pairs between synthetic and runtime right-eye images. These pairs serve as indicators of significant inconsistencies between the two image types.
To quantify the discrepancy, we employ three well-established image similarity metrics: L1 norm, L2 norm, and the Structural Similarity Index Measure (SSIM). Let $I_{syn}$ represent the synthetic right-eye image and $I_{run}$ denote the runtime right-eye image. The three measures are defined as:
\begin{align}
L1(I_{syn}, I_{run}) & = \sum_{i} |I_{syn}[i] - I_{run}[i]| \\
L2(I_{syn}, I_{run}) & = \sqrt{\sum_{i} (I_{syn}[i] - I_{run}[i])^2} \\
SSIM(I_{syn}, I_{run}) & = \frac{(2 \mu_{I_{syn}} \mu_{I_{run}} + c_1) (2 \sigma_{I_{syn}I_{run}} + c_2)}{(\mu_{I_{syn}}^2 + \mu_{I_{run}}^2 + c_1)(\sigma_{I_{syn}}^2 + \sigma_{I_{run}}^2 + c_2)}
\end{align}

Where $\mu$ denotes the average, $\sigma^2$ is the variance, $\sigma_{I_{syn}I_{run}}$ is the covariance of $I_{syn}$ and $I_{run}$, and $c_1, c_2$ are constants to stabilize the division with a weak denominator.

\subsubsection{Inconsistency Detector}
\label{subsubsec:inconsistency-detector}

Introducing hyperparameters \( \alpha \), \( \beta \), and \( \gamma \), the weighted aggregate discrepancy is obtained by summing these measures:
\begin{equation}
D(I_{syn}, I_{run}) = \alpha \cdot L1(I_{syn}, I_{run}) + \beta \cdot L2(I_{syn}, I_{run}) + \gamma \cdot (1 - SSIM(I_{syn}, I_{run}))
\end{equation}

The hyperparameters \( \alpha \), \( \beta \), and \( \gamma \) can be fine-tuned to reflect the importance of each metric in the context of the application. 
\tool uses Isolation Forest~\cite{liu2008isolation} to perform the inconsistency detection based on $D(I_{syn}, I_{run})$.
During detection, \tool constructs numerous random binary decision trees as isolation trees. Each tree is constructed by recursively partitioning the data based on a randomly selected feature and a random split value between the maximum and minimum values of the selected feature.
The anomalies are isolated early, meaning they are closer to the root of the tree, as they are distinct and inhabit less dense regions. Therefore, they have shorter paths on average in these trees compared to normal data points. The length of this path, averaged over multiple trees, forms the anomaly score of a data point.
There exist two key hyperparameters during detection, (1) how many trees are constructed to form the Isolation Forest (\textit{n\_estimators}), and (2) the proportion of outliers in the dataset (\textit{contamination}), which is used to define the threshold for separating outliers from normal observations.
This methodology ensures a structured and comprehensive approach to identifying inconsistencies.
\section{Experiment Design}
\label{sec:evaluation}

\subsection{Research Questions}
Our experiment is designed to answer the following three research questions:
\begin{itemize}[leftmargin=*, topsep=2pt, itemsep=2pt]
        \item \textbf{RQ1 (Stereo-Mapping Generation):} How effective is our proposed \aemodelFull in generating stereo images, i.e., generating right-eye images from left-eye images of VR apps?	
        \item \textbf{RQ2 (Detecting User-Reported \issuesCap):} How effective is our proposed \tool in detecting user-reported \issues? 
        \item \textbf{RQ3 (Detecting Wild \issuesCap):} How well does \tool perform in detecting \issues in wild real-world VR apps?
\end{itemize}

\subsection{Dataset Construction}
\label{subsec:dataset-construction}

Since there are no existing stereoscopic datasets for VR apps. The three experimental datasets are entirely collected by ourselves from the following sources.

The first dataset is automatically collected on 288 real-world VR apps. 
Based on the interaction simulation and automated testing framework for spatial computing extended reality applications proposed by Li et al.~\cite{paper:openvr-testing-infrastructure}, we instantiate ten automated testing agents that simulate HTC VIVE device models and perform automated app interactions on ten Windows PCs.
The interaction actions are generated randomly.
During the automated execution, the testing agents open the \textit{VR View} of two eyes (side-by-side) on the PC and capture screenshots regularly.
The 288 VR apps are sampled from 4,215 free \textit{VR Only} apps available on Steam~\cite{website:steam-app-store-vr}, which is one of the most popular VR app stores currently~\cite{paper:vr-software-quality}.
To ensure the diversity and representativeness of our dataset, we employ a stratified random sampling strategy, selecting Virtual Reality (VR) apps based on their associated categories (tags) within the Steam platform. We initially select one random application from each category to capture the broadest possible scope of VR experiences. Following this initial sampling, we proceed to randomly sample an additional 2\% of the apps within each category. 
At last, we collect \numberOfAllScreenshotsManuallyCollected stereoscopic screenshots as shown in Figure~\ref{subfig:left-right-mapping}.
Table~\ref{table:issue-statistics-datasets} demonstrates the statistics among manifestation categories in Section~\ref{sec:category}.

The second dataset is composed of screenshots of real-world bug reports collected during the empirical analysis introduced in Section~\ref{sec:mc-study}.
Removing the user-edited screenshots with coverings (e.g., texts), screenshots with only monocular views, partial screenshots, etc. leaving us a buggy dataset of \numberOfRealTestPairs stereoscopic screenshots (or left-right image pairs).

For ease of quantitative evaluation, we construct the third dataset. It is a labeled dataset.
We randomly sample a subset from the first dataset, comprising 4,000 VR screenshots. We hire three VR users to rigorously analyze and label these images. All of them are equipped with real VR devices (device model: Pico 4 Pro), using head-mounted displays to view the image in the S3D mode. Each user independently examines the screenshots and labels those stereo-inconsistent images that make them feel cybersickness as \issues.
To ensure objectiveness, we only consider screenshots that have been marked by all three users as the final set of \issues, other screenshots are regarded as normal cases.
After manual inspection, we get 237 \issues.
We further classify these \issues into the 14 manifestation categories as introduced in Section~\ref{sec:category}.
Table~\ref{table:issue-statistics-datasets} demonstrates the statistics.

\begin{table}
    \centering
    \caption{Manifestation category statistics of \issues in our datasets}
    \vspace{-1em}
    \resizebox{0.65\columnwidth}{!}{%
    \begin{tabular}{lrr} 
    \toprule
         &   Dataset Collected from Online Bug Reports &  Sampled Screenshot Dataset \\ \midrule
         Object Omission
&  16
& 44\\ 
         Lighting and Shadow Discrepancies
&  11
& 0\\ 
         Object Position Discrepancy
&  9
& 105\\ 
         Shader Absence
&  6
& 0\\ 
         Monocular Blindness 
&  5
& 17\\ 
         Particle and Visual Effect Variations
&  5
& 0\\ 
         Unilateral Object Rendering
&  5
& 7\\ 
         Material or Texture Mismatch
&  5
& 0\\ 
         Post-Processing Inconsistency
&  5
& 0\\ 
 Level of Detail Inconsistency
& 4
&0\\ 
 Object Warping
& 4
&21\\ 
 View Misalignment
& 3&0\\ 
 Partial Object Rendering
& 1
&6\\ 
 Warped Views 
& 1
&0\\ 
 Asymmetric Viewing Angles& 0&37\\
 Other& 2&0\\ \midrule
 Total & 82&237
\\
 \bottomrule
    \end{tabular}
    \label{table:issue-statistics-datasets}
    }
    \vspace{-1.7em}
\end{table}

\subsection{Implementation Details and Experimental Setup}
\label{subsec:impl-and-exp-setup}

\begin{figure}[t!] 
	\centering 
	\includegraphics[width=0.4\columnwidth]{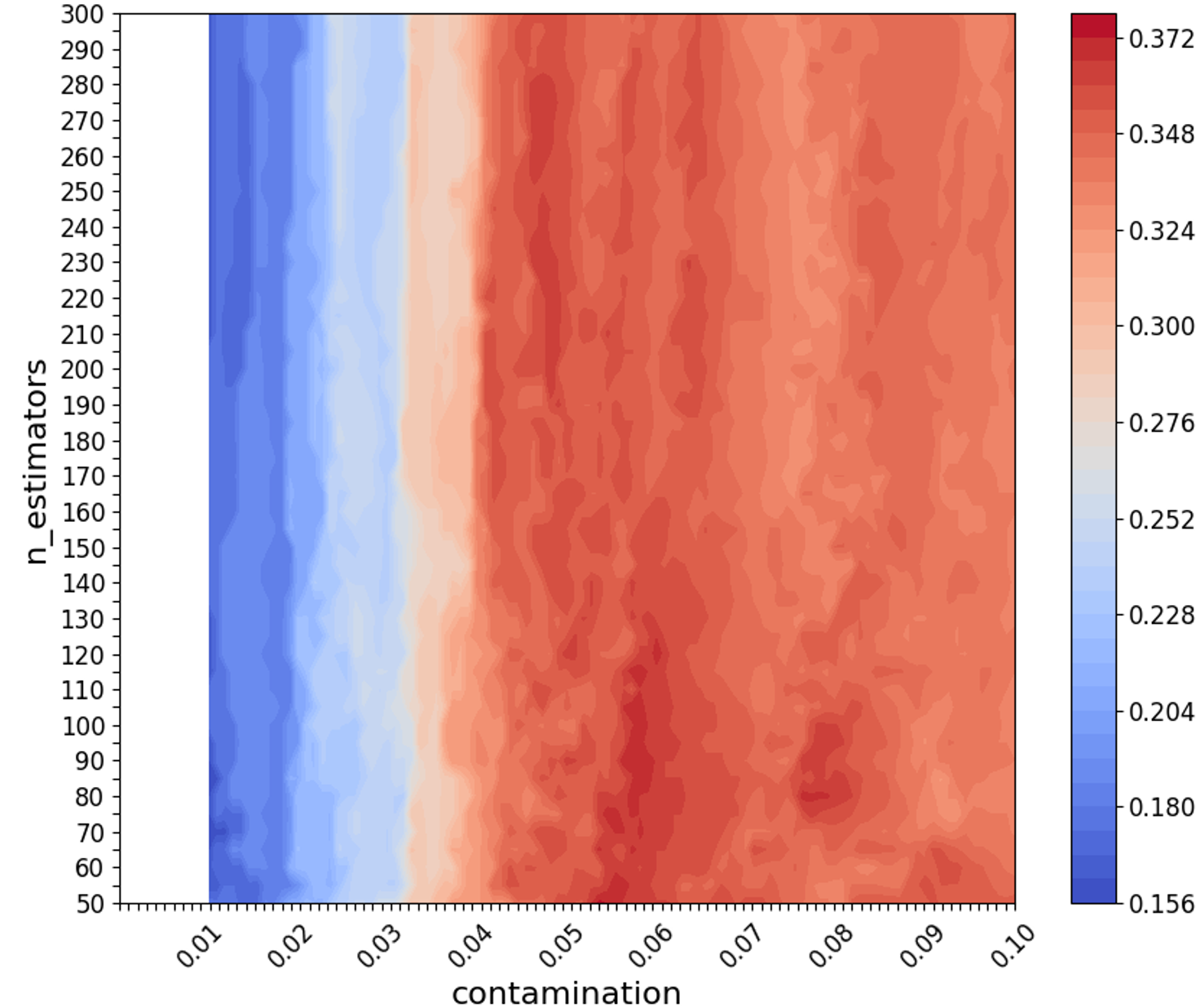} 
 \vspace{-1em}
	\caption{Experiment results of hyperparameter settings of inconsistency detector (F-1 score)}
	\label{fig:hyperparameter-tuning}
  \vspace{-1.5em}
\end{figure}

We implement the \aemodelShort in Python using the PyTorch framework \cite{paszke2019pytorch}.
The U-Net architecture of the generator is with 256 feature maps.
The model is trained by Adam optimizer \cite{kingma2014adam} over batches of 64 input samples with an initial learning rate of 0.001.
The activation functions we use are ReLU (for the generator) and LeakyReLU (for the discriminator) \cite{xu2015empirical}.
We resize the binocular images to 1024 * 576. During training, we scale the input size of monocular images to 512 pixels in width before performing a random crop to a 512 * 512 dimension.
To ensure normalization at each layer of the network, we adopt instance normalization.
For the model used in-depth estimation, we directly use the pre-trained model from the MiDaS v3.1 model suite\cite{ranftl2020towards}.
For experiments, we split the first screenshot dataset into training, validation, and testing sets with a ratio of 90\%, 5\%, 5\%, resulting in 154,566, 8,587, and 8,587 screenshots in each group.
Due to the limit of computation resources, we randomly sample a subset of the training set containing 20,000 screenshots, instead of training \aemodelShort on the whole training set.
Experiments for RQ1 use the aforementioned subset (20,000) for training, the original validation set (8,587) for validation, and the original test set (8,587) for testing.
\textblue{
To detect user-reported \issues, we need to put them into a complete testing set with both positive samples and negative samples.
To make the results align with real-world scenarios, we align the SVI issue ratio of the whole testing set for RQ2 with the natural SVI issue ratio in real-world data distribution.
As demonstrated in Table 1, our manual analysis results show that the natural ratio, i.e., ``\issues:the whole set'' is close to ``\textit{237:4,000}''.
We sample the appropriate number (i.e., 1,302) of the manually-verified normal data (see Section~\ref{subsec:dataset-construction}) and mix them with user-reported \issues (total number: 1,384), forming a testing set under the real-world ratio to evaluate RQ2.
For RQ3, the experiments are conducted on the manual-labeled dataset (see Section~\ref{subsec:dataset-construction}).
}
\textred{
There are several key hyperparameters for Inconsistency Detector (see Section~\ref{subsubsec:inconsistency-detector}).
For \textit{contamination} (i.e., 1 - threshold) and \textit{n\_estimators}, we randomly split the manual-labeled dataset into a ``training set'' and a test set (6:4) for parameter tuning.
Guided by literature, we conduct a grid search using the metrics of F-1 score for detecting SVI issues for 100 potential \textit{contamination} values from 0.01 to 0.1 evenly spaced, and \textit{n\_estimators} from 50 to 300 with a step of 5.
As shown in Figure~\ref{fig:hyperparameter-tuning}, \tool reaches best results on the F-1 score of detecting SVI issues when \textit{contamination} equals 0.058 (i.e., threshold equals 0.942), \textit{n\_estimators} equals 110.
Subsequent experiments on RQ2 and RQ3 are conducted using these hyperparameters.
Other parameters use default or recommended values, e.g., we set \( \alpha \), \( \beta \), and \( \gamma \) of Inconsistency Detector as 1. 
}

\subsection{Baselines}

\subsubsection{For Stereo Generation of \aemodelShortCap}
\label{sec:bl-sg}
\textred{To assess the effectiveness of our proposed \aemodelShort, we compare it against five well-established baselines that are widely used: three in the image generation task and two in the 2D to 3D image conversion task.
All five baselines use default or recommended configurations and hyperparameters from the original papers.}

\begin{itemize}[leftmargin=*, topsep=2pt, itemsep=2pt]
	\item \textbf{ResNet-Based Autoencoder~\cite{wickramasinghe2021resnet}.} A simple yet powerful neural network structure for image reconstruction.
	\item \textbf{U-Net}~\cite{paper:conf/miccai/RonnebergerFB15}. A convolutional network, especially efficient for image segmentation.
	\item \textbf{CycleGAN}~\cite{paper:conf/iccv/ZhuPIE17}. A generative adversarial network, well-known for image-to-image translation.
  	\item \textred{\textbf{Deep3D}~\cite{paper:conf/eccv/XieGF16}. A widely-used 2D to stereoscopic 3D image conversion approach using deep convolutional neural network (CNN).}
 	\item \textred{\textbf{DIBR}~\cite{paper:journals/entropy/Hachaj23}. The state-of-the-art 2D to 3D image conversion approach, utilizes a deep CNN alongside a fast inpaint algorithm.}
\end{itemize}
\textred{We also use a variant of our model without depth guidance as a baseline:}
\begin{itemize}[leftmargin=*, topsep=2pt, itemsep=2pt]
        \item \textbf{\aemodelShortCap without Depth Guidance (\aemodelNoDepth).} Another version of our \aemodelShortCap with depth guidance removed. The whole framework without depth guidance is referred to as \toolNoDepth.
\end{itemize}

\subsubsection{For \issueCapSingular Detection of \tool}

        Use the stereo generation method in Section~\ref{sec:bl-sg}, our distance calculator, and our inconsistency detector for issue detection.
        \textred{The stereo generation methods include ResNet-Based Autoencoder, U-Net, CycleGAN, Deep3D, DIBR, and \aemodelNoDepth.}

\subsection{Evaluation Metrics}

\subsubsection{For Stereo Generation of \aemodelShortCap}
We adopt three distinct evaluation metrics in experiments.
\begin{itemize}[leftmargin=*, topsep=2pt, itemsep=2pt]
	\item \textbf{L1.} The mean absolute error between the synthesized right-eye image and the actual one.
	\item \textbf{L2.} The mean square error between the synthesized right-eye image and the actual right-eye image.
 Smaller L1 and L2 values indicate better performance in terms of pixel-level accuracy.
        \item \textbf{SSIM (Structural Similarity Index Measure).}  Unlike traditional metrics like MSE (Mean Squared Error), SSIM considers changes in structural information, luminance, and contrast. It provides a more perceptual assessment, aiming to capture the human visual system's judgment of quality. Higher SSIM values indicate better performance of \aemodelShort.
\end{itemize}

\subsubsection{For \issueCapSingular Detection of \tool (RQ3)}

A comprehensive set of evaluation metrics was chosen to assess \tool. These metrics enable both a granular insight into the detection of each class and an overall performance understanding. Specifically, we utilize:

\begin{itemize}[leftmargin=*, topsep=2pt, itemsep=2pt]
    \item \textbf{Precision (-1)}: Quantifies the proportion of true \issues among the instances labeled as anomalies by the model. It is defined as:
    \begin{equation}
    P_{-1} = \frac{TP_{-1}}{TP_{-1} + FP_{-1}}
    \end{equation}
    
    \item \textbf{Recall (-1)}: Measures the proportion of actual \issues that were correctly detected. 
    \begin{equation}
    R_{-1} = \frac{TP_{-1}}{TP_{-1} + FN_{-1}}
    \end{equation}
    
    \item \textbf{F1-score (-1)}: The harmonic mean of Precision and Recall for \issues, offering a balance between the two. It is computed as:
    \begin{equation}
    F1_{-1} = 2 \times \frac{P_{-1} \times R_{-1}}{P_{-1} + R_{-1}}
    \end{equation}

    \item \textbf{Precision (1)}: Analogous to Precision (-1), but for the normal class.
    
    \item \textbf{Recall (1)}: Similar to Recall (-1), but captures the detection rate of the normal instances.
    
    \item \textbf{F1-score (1)}: Represents the balance between Precision and Recall for the normal instances.
    
    \item \textbf{Accuracy}: Highlights the proportion of all instances (both \issues and normal instances) that are correctly classified. It's given by:
    \begin{equation}
    \text{Accuracy} = \frac{TP_{-1} + TP_1}{TP_{-1} + TP_1 + FP_{-1} + FP_1 + FN_{-1} + FN_1}
    \end{equation}

    \item \textbf{Macro avg}: The average score across both classes, treating them equally irrespective of their imbalance.
    
    \item \textbf{Weighted avg}: The average score across both classes, but weighted by the number of instances in each class to account for potential class imbalances.
\end{itemize}

Where \(TP_{-1}\), \(FP_{-1}\), and \(FN_{-1}\) denote the true positives, false positives, and false negatives for the anomaly class (-1), respectively, and similarly for the non-anomalous class (1).

For RQ2, since we don't have ground truths for the test set, we can only calculate the detected \issues, undetected \issues and recall on the subset of bug report screenshot.

\section{Results and Analysis}
\label{sec:results}

\subsection{Performance of Stereo Generation (RQ1)}

\begin{table}[t]
    \centering
    \caption{Performance of \aemodelFull (\aemodel) compared with baselines}
    \vspace{-1em}
    \label{table:111}
    \resizebox{\columnwidth}{!}{%
    \begin{tabular}{
        l|rrrrrr|rrrr|rrrr
    }
        \toprule
        Approach & \multicolumn{2}{c}{Autoencoder} & \multicolumn{2}{c}{U-Net} & \multicolumn{2}{c}{CycleGAN} & \multicolumn{2}{c}{\textred{Deep3D}} & \multicolumn{2}{c}{\textred{DIBR}} & \multicolumn{2}{c}{\textbf{\aemodelNoDepth}} & \multicolumn{2}{c}{\textbf{\aemodel}} \\
        \cmidrule(lr){1-1} \cmidrule(lr){2-3} \cmidrule(r){4-5} \cmidrule(r){6-7} \cmidrule(r){8-9} \cmidrule(r){10-11} \cmidrule(r){12-13} \cmidrule(r){14-15}
        Metrics & {avg} & {std} & {avg} & {std} & {avg} & {std} & {\textred{avg}} & {\textred{std}} & {\textred{avg}} & {\textred{std}} & {avg} & {std} & {avg} & {std} \\
        \midrule
        L1 $\downarrow$ & 0.1737 & 0.1117 & 0.0521 & 0.0356 & 0.0230 & 0.0382 & \textred{0.2966} & \textred{0.2086} & \textred{0.0381} & \textred{0.0465} & 0.0219 & 0.0372 & 0.0204 & 0.0366 \\
        L2 $\downarrow$ & 0.2141 & 0.1138 & 0.0862 & 0.0456 & 0.0553 & 0.0575 & \textred{0.3517} & \textred{0.2069} & \textred{0.0861} & \textred{0.0709} & 0.0555 & 0.0556 & 0.0495 & 0.0529 \\
        SSIM $\uparrow$ & 0.3639 & 0.2534 & 0.5223 & 0.3013 & 0.8287 & 0.1667 & \textred{0.2840} & \textred{0.2409} & \textred{0.7903} & \textred{0.1957} & 0.8417 & 0.1622 & 0.8638 & 0.1455 \\
        \bottomrule
    \end{tabular}%
    }
    \vspace{-1em}
\end{table}

\textred{
To evaluate the effectiveness of our proposed \aemodelFull (\aemodel) of \tool in generating stereo images, we conducted extensive experiments comparing \tool with five baseline approaches: ResNet-Based Autoencoder, U-Net, CycleGAN, Deep3D, and DIBR, along with a variant of our model without depth guidance (i.e., \aemodelNoDepth). The comparison focused on three key metrics: L1 loss, L2 loss, and the SSIM, where lower L1 and L2 losses indicate higher pixel-level accuracy, and higher SSIM scores signify better structural and perceptual quality of the generated images.
}

\textred{
Table~\ref{table:111} presents the performance of \aemodel compared with the baseline models. It is evident from the results that \aemodel achieves the lowest average L1 and L2 losses, at 0.0204 and 0.0495, respectively, and the highest SSIM score, at 0.8638. These results signify that \aemodel is capable of generating right-eye images with superior pixel-level accuracy and structural consistency compared to the baseline methods. Notably, \aemodel outperforms the \aemodelNoDepth variant, demonstrating the importance of depth guidance in enhancing the quality of stereo image generation.
}

\textred{
The superior performance of \aemodel can be attributed to its depth-aware image translation mechanism, which incorporates depth information to accurately model the spatial shift between the left and right-eye images. This depth awareness enables \aemodel to better capture the perspective shifts and occlusions that occur in stereo imaging, resulting in more accurate and consistent right-eye images. The high SSIM score further indicates that \aemodel maintains the structural integrity of the scene, which is crucial for avoiding disorientation and cybersickness in VR apps.
}

\textred{
In contrast, baseline methods such as Deep3D and DIBR, while effective in certain contexts, exhibit higher L1 and L2 losses, indicating less accurate pixel-level image generation. Their lower SSIM scores also suggest that these methods may struggle to preserve the structural and perceptual qualities of the original scenes when generating stereo images.
}

\textred{
The results from our evaluation underscore the effectiveness of \aemodel in addressing the challenge of generating accurate and consistent right-eye images for VR apps. By leveraging depth information, \aemodel significantly enhances the quality of stereo-mapping generation, offering a promising solution to mitigate \issues.
}

\subsection{Performance of \tool: Detecting User-Repoted \issuesCap (RQ2)}

\begin{table}[t!]
\centering
\caption{Performance of \issueSingular detection in each manifestation category, Y denotes detected \issues, N denotes undetected SVI issues\protect\footnotemark}
\vspace{-1em}
\label{table:rq2}
\small
\resizebox{0.45\columnwidth}{!}{%
\begin{tabular}{lrrrr}
\toprule
\textbf{\textblue{Category}} &
  \multicolumn{1}{c}{\textbf{\textblue{Y}}} &
  \multicolumn{1}{c}{\textbf{\textblue{N}}} &
  \multicolumn{1}{c}{\textbf{\textblue{Total}}} &
  \multicolumn{1}{c}{\textbf{\textblue{Recall}}} \\ \midrule
\textblue{Object Omission}                       & \textblue{14}& \textblue{2}& \textblue{16}& \textblue{87.5\%}\\
\textblue{Lighting and Shadow Discrepancies}     & \textblue{10}& \textblue{2}& \textblue{11}                       & \textblue{90.9\%}\\
\textblue{Object Position Discrepancy}           & \textblue{5}& \textblue{4}& \textblue{9}                        & \textblue{55.6\%}\\
\textblue{Shader Absence}                        & \textblue{3} & \textblue{3} & \textblue{6}                        & \textblue{50.0\%}  \\
\textblue{Unilateral Object Rendering}           & \textblue{4}& \textblue{1}& \textblue{5}                        & \textblue{80.0\%}  \\
\textblue{Particle and Visual Effect Variations} & \textblue{5}& \textblue{0}& \textblue{5}                        & \textblue{100.0\%}\\
\textblue{Material or Texture Mismatch}          & \textblue{4} & \textblue{1} & \textblue{5}                        & \textblue{80.0\%}  \\
\textblue{Monocular Blindness}                   & \textblue{5} & \textblue{0} & \textblue{5}                        & \textblue{100.0\%} \\
\textblue{Post-Processing Inconsistency}         & \textblue{2} & \textblue{3} & \textblue{5}                        & \textblue{40.0\%}  \\
\textblue{Object Warping}                        & \textblue{4} & \textblue{0} & \textblue{4}                        & \textblue{100.0\%}  \\
\textblue{Level of Detail Inconsistency}         & \textblue{3} & \textblue{1} & \textblue{4}                        & \textblue{75.0\%}  \\
\textblue{View Misalignment}                     & \textblue{3} & \textblue{0} & \textblue{3}                        & \textblue{100.0\%}  \\
\textblue{Warped Views}                          & \textblue{0} & \textblue{1} & \textblue{1}                        & \textblue{0.0\%} \\
\textblue{Partial Object Rendering}              & \textblue{1} & \textblue{0} & \textblue{1}                        & \textblue{100.0\%} \\
\textblue{Asymmetric Viewing Angles}             & \textblue{0} & \textblue{0} & \textblue{0}                        & \textblue{-} \\ 
\textblue{Other}                                 & \textblue{2} & \textblue{0} & \textblue{2}                        & \textblue{100.0\%} \\ \midrule
\textblue{Total}                                 & \textblue{65}& \textblue{17}& \textblue{82} 
      & \textblue{79.3\%}\\

\bottomrule
\end{tabular}%
}
\vspace{-1em}
\end{table}

\footnotetext{Some samples have multiple manifestations and have been categorized in multiple classes.}

\textblue{
This section evaluates the capability of \tool in accurately identifying user-reported SVI issues across various manifestation categories (see Section~\ref{subsec:impl-and-exp-setup}). 
}
\textblue{
Table~\ref{table:rq2} summarizes the detection performance of \tool across each SVI category, presenting the number of issues detected (Y), undetected (N), the total number of issues, and the recall rate. The overall recall rate of \tool stands at 79.3\%, indicating a high degree of effectiveness in identifying SVI issues across most categories. Notably, \tool achieved a 100\% recall rate in several challenging categories, including Particle and Visual Effect Variations, Monocular Blindness, Object Warping, and View Misalignment, showcasing its robust detection capabilities even for semantically-complex \issues. The comprehensive coverage across diverse manifestations of SVI issues demonstrates the adaptability and generalizability of \tool.
}

\textblue{
The analysis also reveals specific areas where \tool could be further optimized, offering directions for future research and development.
Categories such as Post-Processing Inconsistency and Shader Absence reflected not as high recall rates as other issue categories, suggesting areas for further refinement.
This indicates the challenge of identifying subtle or semantic-driven inconsistencies, where the spatial or contextual cues are less pronounced. These findings open research opportunities for integrating more nuanced detection mechanisms or leveraging additional contextual information to improve detection accuracy in these areas.
}
\textblue{
Our results confirm that \tool effectively identifies a wide range of SVI issues in VR apps, marking a significant advancement in automated SVI detection tools. By achieving high recall rates across various manifestations, \tool provides a valuable resource for developers and quality assurance teams to proactively address and rectify SVI issues, thereby enhancing the overall quality and safety of VR environments.
}

\subsection{Performance of Detecting Wild \issuesCap (RQ3)}

\begin{table}[t!]
\centering
\caption{Performance of \issueSingular detection compared with baselines, -1 represents \issues, 1 represents normal cases}
\vspace{-1em}
\label{table:333}
\resizebox{0.69\columnwidth}{!}{%
\begin{tabular}{l|ccc|cc|cc}
\toprule
Metrics & Autoencoder & U-Net & CycleGAN & Deep3D & DIBR & \textbf{\toolNoDepth} & \textbf{\tool} \\
\midrule
Precision (-1) & 5.19 & 32.03 & 31.60 & 2.16 & 17.75 & 34.63 & \cellcolor{gray!40}38.53 \\
Recall (-1)    & 5.04 & 31.09 & 30.67 & 2.10 & 17.23 & 33.61 & \cellcolor{gray!40}37.39 \\
F1-score (-1)  & 5.12 & 31.56 & 31.13 & 2.13 & 17.48 & 34.12 & \cellcolor{gray!40}37.95 \\
\midrule
Precision (1)  & 93.99 & 95.64 & 95.61 & 93.80 & 94.76 & 95.80 & \cellcolor{gray!40}96.04 \\
Recall (1)     & 94.16 & 95.82 & 95.79 & 93.98 & 94.94 & 95.98 & \cellcolor{gray!40}96.22 \\
F1-score (1)   & 94.08 & 95.73 & 95.70 & 93.89 & 94.85 & 95.89 & \cellcolor{gray!40}96.13 \\
\midrule
Accuracy       & 88.85 & 91.95 & 91.90 & 88.50 & 90.30 & 92.26 & \cellcolor{gray!40}92.71 \\
Macro avg      & 49.59 & 63.84 & 63.61 & 47.98 & 56.25 & 65.21 & \cellcolor{gray!40}67.28 \\
Weighted avg   & 88.69 & 91.84 & 91.79 & 88.34 & 90.17 & 92.20 & \cellcolor{gray!40}92.61 \\
\bottomrule
\end{tabular}%
}
\vspace{-1em}
\end{table}

\textred{
In this section, we delve into the performance of \tool in wild real-world VR scenarios. This assessment is crucial for understanding the practical applicability and robustness of \tool when faced with diverse and complex (industrial-setting) real-world VR apps.
We compare \tool against five baselines: Autoencoder, U-Net, CycleGAN, Deep3D, and DIBR, along with a variant of our framework, \toolNoDepth, which lacks depth guidance. Performance metrics include precision, recall, F1-score, and overall accuracy for detected SVI issues (-1) and normal cases (1).
}

\textred{
Table~\ref{table:333} presents the detection performance of \tool and baseline methods. Notably, \tool demonstrates superior performance across all metrics, particularly in the detection of SVI issues, with a precision of 38.53\%, recall of 37.39\%, and F1-score of 37.95\%. These results are highlighted in comparison to normal case detection, where \tool also leads with precision, recall, and F1-score of 96.04\%, 96.22\%, and 96.13\%, respectively. Overall, \tool achieved an accuracy of 92.71\%, outperforming all baselines and evidencing its effectiveness in real-world scenarios.
\tool's leading performance in detecting SVI issues underscores its capability to accurately identify inconsistencies in complex VR environments. The improvement over baselines, especially in detecting SVI issues, is significant, considering the challenges posed by real-world apps, such as diverse visual effects, dynamic content, and varying levels of detail. The depth-aware mechanism within \tool is pivotal for this success, enabling it to discern subtle discrepancies that are often missed by other approaches.
The superior recall and precision in normal case detection further illustrate \tool's robustness, minimizing false positives and ensuring reliable identification of non-issues. This balance is crucial for practical apps, where over-reporting can burden developers with unnecessary reviews, and under-reporting can leave critical issues undetected.
}

\textred{
The evaluation results in the wild reinforce the potential of \tool as a valuable tool for developers and testers of VR apps. By offering high accuracy and robust performance across a spectrum of real-world scenarios, \tool stands out as a promising solution for enhancing the quality assurance process in VR development.
Moreover, the results highlight the importance of incorporating depth information in SVI detection tools, opening avenues for future research to explore more sophisticated depth-aware approaches for even greater accuracy.
}

\subsection{\textblue{Statistical Significance Analysis}}

\textblue{
We conducted Mann–Whitney U test to analyze whether there exist statistically significant differences between the results achieved by our approach and those of baselines. This non-parametric test is particularly apt for analyzing data that do not necessarily follow a normal distribution, making it a robust choice for our evaluation. Specifically, the test is conducted under the distance calculation results between the generated right-eye images and actual right-eye images. 
}

\textblue{
As shown in Table~\ref{table:rq1-p-value}, these results collectively demonstrate the statistical significance of our method's performance when compared with baselines. The consistently low (or near-zero) p-values across different comparisons unequivocally suggest that our approach is significantly better than baselines in terms of detecting \issues.
}

\begin{table}[t!]
\centering
\caption{\textblue{Results of statistical significance analysis}}
\vspace{-1em}
\label{table:rq1-p-value}
\resizebox{0.8\columnwidth}{!}{%
\begin{tabular}{@{}l|ccc|cc|c@{}}
\toprule
\multicolumn{1}{l}{\begin{tabular}[c]{@{}l@{}} \textbf{Analyzed Between Our Approach} \\ \textbf{and Which Baseline} \end{tabular}} & \textblue{Autoencoder} & \textblue{U-Net} & \textblue{CycleGAN} & \textblue{Deep3D} & \textblue{DIBR} & \textblue{\toolNoDepth} \\ 
\midrule
\textblue{P-Value} & \textblue{0.0000} & \textblue{0.0000} & \textblue{7.8358e-19} & \textblue{0.0000} & \textblue{4.0191e-235} & \textblue{1.0528e-05} \\
\bottomrule
\end{tabular}
}
\vspace{-1em}
\end{table}

\section{Threats to Validity}

\textbf{Internal Validity.} \tool partially depends on deep learning models, specifically conditional GAN models. These models inherently possess a risk of overfitting, which could lead to high performance on our dataset but limited generalization to unseen data. To mitigate this threat, we employ a rigorous training protocol, involving validation sets and early stopping mechanisms.

\textbf{External Validity.} While our empirical study incorporates a wide range of VR apps, it is conceivable that certain types of apps are underrepresented. This implies that our findings and the resulting deep learning model may not apply universally to all VR apps. Furthermore, our research is grounded in the current state of VR technology. Future advancements in VR hardware and software may necessitate adaptations in our methodology.

\textbf{Construct Validity.} We identify SVI issues based on a discrepancy metric between synthetic and actual right-eye images. While this discrepancy measure is intuitive and has demonstrated utility in our work, it may not capture all possible manifestations of SVI issues. Certain types of glitches might lead to negligible discrepancy according to our measure, yet still induce cybersickness in users. Furthermore, both the empirical study and data collection of one of our datasets rely on the analysis of bug reports from online forums. The inherent subjectivity and potential inaccuracies of these reports present a risk to the fidelity of our empirical study findings and the collected dataset.
\section{Related Work}
\label{sec:related-work}

\subsection{Studies on VR Applications}
Several empirical studies have been conducted to explore and understand VR apps.
Rodriguez and Wang~\cite{paper:os-vr-rodriguez17} studied the growing trends, popular topics, and common file structures of open-source VR projects.
Adams et al.~\cite{paper:sec-interview-adams18} conducted interviews with VR developers and users to understand their concerns about security and privacy.
Li et al.~\cite{paper:issre-webxr-empirical} performed an empirical study on web-based extended reality bugs to understand their symptoms, root causes, and uniqueness.
Nusrat et al.~\cite{paper:vr-performance-optimization} analyzed the optimization commits in open-source Unity-based VR projects to better understand VR performance optimization.
Epp et al.~\cite{paper:vrgame-empirical-Epp21} studied the characteristics of VR games on Steam and players' complaints about these VR games.
\textred{Rzig et al.~\cite{paper:vr-testing-empirical} conducted a quantitative and qualitative empirical study to analyze existing testing practices of open-source Unity-Based VR apps.}
Recently, Li et al.~\cite{paper:vr-software-quality} conducted a large-scale empirical study to model the software quality of VR apps from users' perspectives. 
Based on the findings of VR software quality, they discussed insightful implications of VR quality assurance for both developers and researchers.

\subsection{\textred{Virtual / Augmented Reality Testing}}
\textred{
In the realm of Extended Reality (XR, including VR/AR/MR) testing, significant strides have been made to address the unique challenges posed by these technologies~\cite{paper:vr-testing-empirical, paper:conf/icse/Wang22, paper:conf/kbse/WangRM23, paper:conf/hci/GilFRCQ20, paper:journals/stvr/SouzaND18, paper:conf/kbse/RafiZW22}. 
For VR testing, Wang~\cite{paper:conf/icse/Wang22} proposed a white-box framework, VRTest, to automate the testing of VR scenes by extracting scene information and manipulating the user camera with predefined testing strategies.
Further refining the approach to VR testing, Wang, Rafi, and Meng~\cite{paper:conf/kbse/WangRM23} developed VRGuide. VRGuide employs a computational geometry method, Cut Extension, to optimize camera routes for comprehensive object interaction coverage in VR scenes.
Gil et al.~\cite{paper:conf/hci/GilFRCQ20} proposed Youkai, a framework tailored for Android-based VR apps. Youkai's capabilities in object detection, camera position adjustment, and support for six Degrees of Freedom scenarios offer promising preliminary results for VR unit testing.
Souza, Nunes, and Dias~\cite{paper:journals/stvr/SouzaND18} introduced VR-ReST,
which facilitates the specification of requirements through a semi-formal language and generates test data from these specifications.
As for AR testing, Rafi et al.~\cite{paper:conf/kbse/RafiZW22} proposed PredART to predict human ratings of virtual object placements in AR scenarios and further detect placement issues. Using automatic screenshot sampling, crowd-sourcing, and a hybrid neural network for image regression, PredART achieves effective results in placement issue detection.
Given the lack of user-side end-to-end dynamic XR testing infrastructure, recently, Li et al.~\cite{paper:openvr-testing-infrastructure} proposed an interaction simulation and automated black-box testing framework for spatial computing XR apps.
}

\textred{
Our work, \tool, uniquely contributes to the VR/AR testing domain by focusing on the detection of \issues without the need for labeled data or access to internal application code. Unlike the methods mentioned above, which primarily target general testing challenges in VR/AR environments, \tool specifically addresses the nuanced issue of SVI through a novel unsupervised, black-box testing framework. Our depth-aware left-right-eye image translator, integral to \tool, signifies a groundbreaking step towards understanding and rectifying SVI issues in VR apps, pushing the boundaries of automated VR testing further.
}

\subsection{Automated GUI Test Oracle}
\label{subsec:related-work-gui-oracle}
Recently, in order to address the limitations of regular test oracles in detecting GUI glitches, researchers have started exploring automated GUI test oracles that \textblue{employ deep learning (DL) techniques for identifying abnormal GUI states in mobile and web apps~\cite{paper:owl-eyes, paper:glib, paper:seenomaly, paper:html5-canvas-test, paper:conf/kbse/RafiZW22}. }
These studies typically model GUI glitch detection as a classification problem and use different methods to perform buggy-side training data augmentation, generating screen captures with GUI glitches by either modifying normal GUI screenshots or injecting bugs into the application code and capturing the resulting screens. 
These data augmentation methods supply sufficient training data for deep learning classifiers to effectively detect faults. 
Such approaches yield promising results in their corresponding scenarios, significantly improving the efficiency of GUI testing.

Liu et al. proposed OwlEye~\cite{paper:owl-eyes}, a deep learning-based approach to detect UI display issues such as text overlap, blurred screens, and missing images. The technique utilizes visual information from GUI screenshots.
Chen et al. introduced GLIB~\cite{paper:glib}, a technique designed for graphically-rich apps like games. GLIB detects non-crashing bugs, such as graphical glitches, using a code-based data augmentation technique. 
Zhao et al. presented Seenomaly~\cite{paper:seenomaly}, a vision-based linting approach for GUI animations. Seenomaly uses an unsupervised, computer-vision-based adversarial autoencoder to group similar GUI animations, even when lacking sufficient labeled data for training. This approach aids in linting GUI animations against design-don't guidelines.
\textgreen{
Macklon et al.~\cite{paper:html5-canvas-test} developed a technique for automatically detecting visual bugs in HTML5 canvas games by leveraging an internal representation of objects on the canvas. The method decomposes snapshot images into object images and compares them to oracle assets using four similarity metrics. 
}

\subsection{\textred{2D-to-3D Image Conversion}}

\textred{
There exist multiple works in the computer vision domain that are related to 2D-to-3D image conversion.
Deep3D by Xie et al.~\cite{paper:conf/eccv/XieGF16} uses deep CNNs for the automatic conversion of 2D videos and images into stereoscopic 3D formats. Unlike traditional methods requiring ground truth depth maps, Deep3D leverages stereo pairs from 3D movies for end-to-end training, showcasing significant advancements in performance and human subject evaluations.
Lee et al.'s~\cite{paper:conf/icip/LeeJKS17} multi-scale DNN approach redefines view synthesis from a single reference view by integrating a spatial transformer module within a unified CNN framework. 
Chen, Yuan, and Bao's DenseNet3D~\cite{paper:conf/ictai/ChenYB19} introduces a novel application of 3D densely connected convolutional networks for automatic 2D-to-3D video conversion, achieving improved results and speed by incorporating 3D convolution layers to grasp the spatiotemporal video characteristics.
The Fused Network proposed by Zhu, Liu, and Wang~\cite{paper:conf/mipr/ZhuLW20} integrates unsupervised depth estimation with DIBR in an end-to-end framework, demonstrating the benefits of leveraging both contextual and geometric information for view synthesis.
Shih et al.'s~\cite{paper:conf/cvpr/ShihSKH20} context-aware layered depth inpainting technique for 3D photography innovatively combines color and depth structure synthesis in occluded regions, offering an improved method for novel view synthesis in everyday scenes.
Recently, Hachaj's~\cite{paper:journals/entropy/Hachaj23} adaptable 2D-to-3D conversion approach utilizes a deep convolutional neural network alongside a fast inpainting algorithm, emphasizing the adaptability and efficiency of converting 2D images to 3D.
}

\textred{
Our work, \tool, distinctively focuses on the detection of \issues in VR apps, a niche yet critical aspect of the 2D-to-3D conversion domain. 
Unlike the aforementioned studies primarily aimed at enhancing the conversion quality or efficiency, 
\tool tackles the challenge of \issues detection through an innovative unsupervised black-box testing framework. 
The novel depth-aware left-right-eye image translator, proposed by us, can generate high-quality synthetic right-eye images for SVI issue detection.
Results in Section~\ref{sec:results} demonstrate that \tool outperforms the state-of-the-art 2D-to-3D image conversion approaches DIBR~\cite{paper:journals/entropy/Hachaj23} and Deep3D~\cite{paper:conf/eccv/XieGF16}.
}
\section{Conclusion}
\label{sec:conclusion}

This paper presents \tool, an unsupervised black-box testing framework, to detect \issuesFull (\issues) in VR apps. Our empirical analysis of 282 real-world \issueSingular bug reports from 15 VR platforms reveals the complexity and diversity of SVI issues, which are challenging for existing pattern-based supervised GUI testing methods.
\tool leverages a \aemodelFull, \aemodel, to generate synthetic right-eye images from left-eye images, framing \issueSingular detection as an anomaly detection problem.
We construct several datasets to verify the effectiveness and usefulness of \tool, including a large-scale dataset of over 171K VR stereo image screenshots collected from 288 real-world VR apps.
Extensive evaluations demonstrate that \tool effectively identifies SVI issues in both user-reported \issues and wild VR scenarios, outperforming baselines in pixel-level accuracy, structural consistency, and \issueSingular detection accuracy.
We believe our research significantly advances the quality assurance of VR apps, enhancing user experience and safety. 
Our future work will focus on 
refining and enhancing the framework to further improve robustness and applicability.

\section{Data Availability}
We release our datasets and \tool, at \href{https://sites.google.com/view/stereoid}{\color{blue}{https://sites.google.com/view/stereoid}}.

\section*{Acknowledgement}
We would like to thank the anonymous reviewers for their helpful feedback. The work described in this paper was supported by the National Natural Science Foundation of China (Grant No. 62372219), the Research Grants Council of the Hong Kong Special Administrative Region, China (No. CUHK 14206921 of the General Research Fund), the Natural Science Foundation of Guangdong Province (Project No. 2023A1515011959), the Shenzhen International Science and Technology Cooperation Project (No. GJHZ20220913143008015), and the Shenzhen-Hong Kong Joint Funding Project (No. SGDX20230116091246007).

\clearpage

\balance
\bibliographystyle{ACM-Reference-Format}
\bibliography{stereo}


\begin{thebibliography}{74}


\ifx \showCODEN    \undefined \def \showCODEN     #1{\unskip}     \fi
\ifx \showDOI      \undefined \def \showDOI       #1{#1}\fi
\ifx \showISBNx    \undefined \def \showISBNx     #1{\unskip}     \fi
\ifx \showISBNxiii \undefined \def \showISBNxiii  #1{\unskip}     \fi
\ifx \showISSN     \undefined \def \showISSN      #1{\unskip}     \fi
\ifx \showLCCN     \undefined \def \showLCCN      #1{\unskip}     \fi
\ifx \shownote     \undefined \def \shownote      #1{#1}          \fi
\ifx \showarticletitle \undefined \def \showarticletitle #1{#1}   \fi
\ifx \showURL      \undefined \def \showURL       {\relax}        \fi
\providecommand\bibfield[2]{#2}
\providecommand\bibinfo[2]{#2}
\providecommand\natexlab[1]{#1}
\providecommand\showeprint[2][]{arXiv:#2}

\bibitem[web(2021a)]%
        {website:xr-application-xrfilm}
 \bibinfo{year}{2021}\natexlab{a}.
\newblock \bibinfo{title}{{FILM XR}}.
\newblock \bibinfo{howpublished}{\url{https://vrfilmreview.ru/}}.
\newblock


\bibitem[web(2021b)]%
        {website:xr-application-VirtualSkill}
 \bibinfo{year}{2021}\natexlab{b}.
\newblock \bibinfo{title}{{VirtualSkill - Virtual Reality Training}}.
\newblock \bibinfo{howpublished}{\url{https://virtualskill.com/}}.
\newblock


\bibitem[web(2021c)]%
        {website:xr-application-xrgames}
 \bibinfo{year}{2021}\natexlab{c}.
\newblock \bibinfo{title}{{XR Games}}.
\newblock \bibinfo{howpublished}{\url{https://www.xrgames.io/}}.
\newblock


\bibitem[web(2022a)]%
        {website:oculus-app-lab}
 \bibinfo{year}{2022}\natexlab{a}.
\newblock \bibinfo{title}{{Oculus App Lab}}.
\newblock
  \bibinfo{howpublished}{\url{https://developer.oculus.com/blog/introducing-app-lab-a-new-way-to-distribute-oculus-quest-apps/}}.
\newblock


\bibitem[web(2022b)]%
        {website:oculus-app-store}
 \bibinfo{year}{2022}\natexlab{b}.
\newblock \bibinfo{title}{{Oculus App Store}}.
\newblock
  \bibinfo{howpublished}{\url{https://www.oculus.com/experiences/quest/}}.
\newblock


\bibitem[web(2022c)]%
        {website:sidequest}
 \bibinfo{year}{2022}\natexlab{c}.
\newblock \bibinfo{title}{{SideQuest}}.
\newblock \bibinfo{howpublished}{\url{https://sidequestvr.com/}}.
\newblock


\bibitem[web(2023a)]%
        {website:ue-forum}
 \bibinfo{year}{2023}\natexlab{a}.
\newblock \bibinfo{title}{{Epic Developer Community Forums}}.
\newblock \bibinfo{howpublished}{\url{https://forums.unrealengine.com/}}.
\newblock


\bibitem[web(2023b)]%
        {website:github}
 \bibinfo{year}{2023}\natexlab{b}.
\newblock \bibinfo{title}{{GitHub}}.
\newblock \bibinfo{howpublished}{\url{https://github.com/}}.
\newblock


\bibitem[web(2023c)]%
        {website:github-steamvr-linux}
 \bibinfo{year}{2023}\natexlab{c}.
\newblock \bibinfo{title}{{GitHub Repository of
  ValveSoftware/SteamVR-for-Linux}}.
\newblock
  \bibinfo{howpublished}{\url{https://github.com/ValveSoftware/SteamVR-for-Linux}}.
\newblock


\bibitem[web(2023d)]%
        {website:steamvr-home}
 \bibinfo{year}{2023}\natexlab{d}.
\newblock \bibinfo{title}{{Introducing SteamVR Home Beta}}.
\newblock
  \bibinfo{howpublished}{\url{https://steamcommunity.com/games/250820/announcements/detail/1256913672017157095}}.
\newblock


\bibitem[web(2023e)]%
        {website:meta-community-forum}
 \bibinfo{year}{2023}\natexlab{e}.
\newblock \bibinfo{title}{{Meta Community Forums}}.
\newblock \bibinfo{howpublished}{\url{https://communityforums.atmeta.com/}}.
\newblock


\bibitem[web(2023f)]%
        {website:unity-manual-post-processing}
 \bibinfo{year}{2023}\natexlab{f}.
\newblock \bibinfo{title}{{Post-processing and Full-screen Effects}}.
\newblock
  \bibinfo{howpublished}{\url{https://docs.unity3d.com/Manual/PostProcessingOverview.html}}.
\newblock


\bibitem[web(2023g)]%
        {website:unity-post-processing-list}
 \bibinfo{year}{2023}\natexlab{g}.
\newblock \bibinfo{title}{{Post-processing Effects}}.
\newblock
  \bibinfo{howpublished}{\url{https://docs.unity3d.com/Packages/com.unity.render-pipelines.high-definition@12.0/manual/post-processing-effect-list.html}}.
\newblock


\bibitem[web(2023h)]%
        {website:stack-overflow}
 \bibinfo{year}{2023}\natexlab{h}.
\newblock \bibinfo{title}{{Stack Overflow}}.
\newblock \bibinfo{howpublished}{\url{https://stackoverflow.com/}}.
\newblock


\bibitem[web(2023i)]%
        {website:steam-community}
 \bibinfo{year}{2023}\natexlab{i}.
\newblock \bibinfo{title}{{Steam Community}}.
\newblock \bibinfo{howpublished}{\url{https://steamcommunity.com/}}.
\newblock


\bibitem[web(2023j)]%
        {website:unity-discussions}
 \bibinfo{year}{2023}\natexlab{j}.
\newblock \bibinfo{title}{{Unity Discussions}}.
\newblock \bibinfo{howpublished}{\url{https://discussions.unity.com/}}.
\newblock


\bibitem[web(2023k)]%
        {website:unity-forum}
 \bibinfo{year}{2023}\natexlab{k}.
\newblock \bibinfo{title}{{Unity Forum}}.
\newblock \bibinfo{howpublished}{\url{https://forum.unity.com/}}.
\newblock


\bibitem[web(2023l)]%
        {website:right-eye-discrepancies}
 \bibinfo{year}{2023}\natexlab{l}.
\newblock \bibinfo{title}{{Unity Forum: Right Eye Discrepancies}}.
\newblock
  \bibinfo{howpublished}{\url{https://forum.unity.com/threads/right-eye-discrepancies-oculus-urp-obi-fluid-shader.1047632/}}.
\newblock


\bibitem[web(2023m)]%
        {website:unity-issue-tracker}
 \bibinfo{year}{2023}\natexlab{m}.
\newblock \bibinfo{title}{{Unity Issue Tracker}}.
\newblock \bibinfo{howpublished}{\url{https://issuetracker.unity3d.com/}}.
\newblock


\bibitem[web(2023n)]%
        {website:eye-different-lod}
 \bibinfo{year}{2023}\natexlab{n}.
\newblock \bibinfo{title}{{Unreal Engine Forums: Eyes Sometimes Show Different
  LODs in VR}}.
\newblock
  \bibinfo{howpublished}{\url{https://forums.unrealengine.com/t/eyes-sometimes-show-different-lods-in-vr/389839}}.
\newblock


\bibitem[web(2023o)]%
        {website:vive-eyes-displacements}
 \bibinfo{year}{2023}\natexlab{o}.
\newblock \bibinfo{title}{{Unreal Engine Forums: Vive Eyes displacements}}.
\newblock
  \bibinfo{howpublished}{\url{https://forums.unrealengine.com/t/vive-eyes-displacements/101890}}.
\newblock


\bibitem[web(2023p)]%
        {website:ue-issue-tracker}
 \bibinfo{year}{2023}\natexlab{p}.
\newblock \bibinfo{title}{{Unreal Engine Issues and Bug Tracker}}.
\newblock \bibinfo{howpublished}{\url{https://issues.unrealengine.com/}}.
\newblock


\bibitem[web(2023q)]%
        {website:unity-veg}
 \bibinfo{year}{2023}\natexlab{q}.
\newblock \bibinfo{title}{{Visual Effect Graph}}.
\newblock
  \bibinfo{howpublished}{\url{https://docs.unity3d.com/2023.2/Documentation/Manual/VFXGraph.html}}.
\newblock


\bibitem[web(2023r)]%
        {website:vive-forum}
 \bibinfo{year}{2023}\natexlab{r}.
\newblock \bibinfo{title}{{VIVE Forum}}.
\newblock \bibinfo{howpublished}{\url{https://forum.htc.com/}}.
\newblock


\bibitem[web(2023s)]%
        {website:viveport}
 \bibinfo{year}{2023}\natexlab{s}.
\newblock \bibinfo{title}{{VIVEPORT}}.
\newblock \bibinfo{howpublished}{\url{https://www.viveport.com/}}.
\newblock


\bibitem[web(2023t)]%
        {website:steam-app-store-vr}
 \bibinfo{year}{2023}\natexlab{t}.
\newblock \bibinfo{title}{{VR Content on Steam App Store}}.
\newblock
  \bibinfo{howpublished}{\url{https://store.steampowered.com/search/?vrsupport=401}}.
\newblock


\bibitem[web(2023u)]%
        {website:meta-vr-playtest-guide}
 \bibinfo{year}{2023}\natexlab{u}.
\newblock \bibinfo{title}{{VR Playtesting Guide}}.
\newblock
  \bibinfo{howpublished}{\url{https://developer.oculus.com/resources/playtest-guide/}}.
\newblock


\bibitem[Adams et~al\mbox{.}(2018)]%
        {paper:sec-interview-adams18}
\bibfield{author}{\bibinfo{person}{Devon Adams}, \bibinfo{person}{Alseny Bah},
  \bibinfo{person}{Catherine Barwulor}, \bibinfo{person}{Nureli Musaby},
  \bibinfo{person}{Kadeem Pitkin}, {and} \bibinfo{person}{Elissa~M. Redmiles}.}
  \bibinfo{year}{2018}\natexlab{}.
\newblock \showarticletitle{{Ethics Emerging: the Story of Privacy and Security
  Perceptions in Virtual Reality}}. In \bibinfo{booktitle}{\emph{{SOUPS}}}.
  \bibinfo{publisher}{{USENIX} Association}, \bibinfo{pages}{427--442}.
\newblock


\bibitem[Arjovsky et~al\mbox{.}(2017)]%
        {arjovsky2017wasserstein}
\bibfield{author}{\bibinfo{person}{Martin Arjovsky}, \bibinfo{person}{Soumith
  Chintala}, {and} \bibinfo{person}{L{\'e}on Bottou}.}
  \bibinfo{year}{2017}\natexlab{}.
\newblock \showarticletitle{Wasserstein generative adversarial networks}. In
  \bibinfo{booktitle}{\emph{International conference on machine learning}}.
  PMLR, \bibinfo{pages}{214--223}.
\newblock


\bibitem[Chang et~al\mbox{.}(2020)]%
        {paper:vr-sickness-cause-measure-3}
\bibfield{author}{\bibinfo{person}{Eunhee Chang}, \bibinfo{person}{Hyun~Taek
  Kim}, {and} \bibinfo{person}{Byounghyun Yoo}.}
  \bibinfo{year}{2020}\natexlab{}.
\newblock \showarticletitle{{Virtual Reality Sickness: {A} Review of Causes and
  Measurements}}.
\newblock \bibinfo{journal}{\emph{Int. J. Hum. Comput. Interact.}}
  \bibinfo{volume}{36}, \bibinfo{number}{17} (\bibinfo{year}{2020}),
  \bibinfo{pages}{1658--1682}.
\newblock
\urldef\tempurl%
\url{https://doi.org/10.1080/10447318.2020.1778351}
\showDOI{\tempurl}


\bibitem[Chen et~al\mbox{.}(2019)]%
        {paper:conf/ictai/ChenYB19}
\bibfield{author}{\bibinfo{person}{Bei Chen}, \bibinfo{person}{Jiabin Yuan},
  {and} \bibinfo{person}{Xiuping Bao}.} \bibinfo{year}{2019}\natexlab{}.
\newblock \showarticletitle{{Automatic 2D-to-3D Video Conversion using 3D
  Densely Connected Convolutional Networks}}. In \bibinfo{booktitle}{\emph{31st
  {IEEE} International Conference on Tools with Artificial Intelligence,
  {ICTAI} 2019, Portland, OR, USA, November 4-6, 2019}}.
  \bibinfo{publisher}{{IEEE}}, \bibinfo{pages}{361--367}.
\newblock
\urldef\tempurl%
\url{https://doi.org/10.1109/ICTAI.2019.00058}
\showDOI{\tempurl}


\bibitem[Chen et~al\mbox{.}(2021)]%
        {paper:glib}
\bibfield{author}{\bibinfo{person}{Ke Chen}, \bibinfo{person}{Yufei Li},
  \bibinfo{person}{Yingfeng Chen}, \bibinfo{person}{Changjie Fan},
  \bibinfo{person}{Zhipeng Hu}, {and} \bibinfo{person}{Wei Yang}.}
  \bibinfo{year}{2021}\natexlab{}.
\newblock \showarticletitle{{{GLIB:} Towards Automated Test Oracle for
  Graphically-Rich Applications}}. In \bibinfo{booktitle}{\emph{{ESEC/FSE}}}.
  \bibinfo{publisher}{{ACM}}, \bibinfo{pages}{1093--1104}.
\newblock
\urldef\tempurl%
\url{https://doi.org/10.1145/3468264.3468586}
\showDOI{\tempurl}


\bibitem[Creswell and Poth(2016)]%
        {book:open-coding-16}
\bibfield{author}{\bibinfo{person}{John~W Creswell} {and}
  \bibinfo{person}{Cheryl~N Poth}.} \bibinfo{year}{2016}\natexlab{}.
\newblock \bibinfo{booktitle}{\emph{Qualitative inquiry and research design:
  Choosing among five approaches}}.
\newblock \bibinfo{publisher}{Sage publications}.
\newblock


\bibitem[Elmqaddem(2019)]%
        {paper:s3d-2}
\bibfield{author}{\bibinfo{person}{Noureddine Elmqaddem}.}
  \bibinfo{year}{2019}\natexlab{}.
\newblock \showarticletitle{{Augmented Reality and Virtual Reality in
  Education. Myth or Reality?}}
\newblock \bibinfo{journal}{\emph{{Int. J. Emerg. Technol. Learn.}}}
  \bibinfo{volume}{14}, \bibinfo{number}{3} (\bibinfo{year}{2019}),
  \bibinfo{pages}{234--242}.
\newblock
\urldef\tempurl%
\url{https://doi.org/10.3991/ijet.v14i03.9289}
\showDOI{\tempurl}


\bibitem[Epp et~al\mbox{.}(2021)]%
        {paper:vrgame-empirical-Epp21}
\bibfield{author}{\bibinfo{person}{Rain Epp}, \bibinfo{person}{Dayi Lin}, {and}
  \bibinfo{person}{Cor{-}Paul Bezemer}.} \bibinfo{year}{2021}\natexlab{}.
\newblock \showarticletitle{{An Empirical Study of Trends of Popular Virtual
  Reality Games and Their Complaints}}.
\newblock \bibinfo{journal}{\emph{{IEEE} Trans. Games}} \bibinfo{volume}{13},
  \bibinfo{number}{3} (\bibinfo{year}{2021}), \bibinfo{pages}{275--286}.
\newblock
\urldef\tempurl%
\url{https://doi.org/10.1109/TG.2021.3057288}
\showDOI{\tempurl}


\bibitem[Gil et~al\mbox{.}(2020)]%
        {paper:conf/hci/GilFRCQ20}
\bibfield{author}{\bibinfo{person}{Adriano~M. Gil}, \bibinfo{person}{Thiago~S.
  Figueira}, \bibinfo{person}{Elton Ribeiro}, \bibinfo{person}{Afonso~R.
  Costa}, {and} \bibinfo{person}{Pablo Quiroga}.}
  \bibinfo{year}{2020}\natexlab{}.
\newblock \showarticletitle{{Automated Test of {VR} Applications}}. In
  \bibinfo{booktitle}{\emph{{HCI} International 2020 - Late Breaking Posters -
  22nd International Conference, {HCII} 2020, Copenhagen, Denmark, July 19-24,
  2020, Proceedings, Part {II}}} \emph{(\bibinfo{series}{Communications in
  Computer and Information Science}, Vol.~\bibinfo{volume}{1294})}.
  \bibinfo{publisher}{Springer}, \bibinfo{pages}{145--149}.
\newblock
\urldef\tempurl%
\url{https://doi.org/10.1007/978-3-030-60703-6\_18}
\showDOI{\tempurl}


\bibitem[Gulrajani et~al\mbox{.}(2017)]%
        {gulrajani2017improved}
\bibfield{author}{\bibinfo{person}{Ishaan Gulrajani}, \bibinfo{person}{Faruk
  Ahmed}, \bibinfo{person}{Martin Arjovsky}, \bibinfo{person}{Vincent
  Dumoulin}, {and} \bibinfo{person}{Aaron~C Courville}.}
  \bibinfo{year}{2017}\natexlab{}.
\newblock \showarticletitle{Improved training of wasserstein gans}.
\newblock \bibinfo{journal}{\emph{Advances in neural information processing
  systems}}  \bibinfo{volume}{30} (\bibinfo{year}{2017}).
\newblock


\bibitem[Hachaj(2023)]%
        {paper:journals/entropy/Hachaj23}
\bibfield{author}{\bibinfo{person}{Tomasz Hachaj}.}
  \bibinfo{year}{2023}\natexlab{}.
\newblock \showarticletitle{{Adaptable 2D to 3D Stereo Vision Image Conversion
  Based on a Deep Convolutional Neural Network and Fast Inpaint Algorithm}}.
\newblock \bibinfo{journal}{\emph{Entropy}} \bibinfo{volume}{25},
  \bibinfo{number}{8} (\bibinfo{year}{2023}), \bibinfo{pages}{1212}.
\newblock
\urldef\tempurl%
\url{https://doi.org/10.3390/E25081212}
\showDOI{\tempurl}


\bibitem[Isola et~al\mbox{.}(2017)]%
        {isola2017image}
\bibfield{author}{\bibinfo{person}{Phillip Isola}, \bibinfo{person}{Jun-Yan
  Zhu}, \bibinfo{person}{Tinghui Zhou}, {and} \bibinfo{person}{Alexei~A
  Efros}.} \bibinfo{year}{2017}\natexlab{}.
\newblock \showarticletitle{Image-to-image translation with conditional
  adversarial networks}. In \bibinfo{booktitle}{\emph{Proceedings of the IEEE
  conference on computer vision and pattern recognition}}.
  \bibinfo{pages}{1125--1134}.
\newblock


\bibitem[Jeon et~al\mbox{.}(2015)]%
        {paper:conf/cvpr/JeonPCPBTK15}
\bibfield{author}{\bibinfo{person}{Hae{-}Gon Jeon}, \bibinfo{person}{Jaesik
  Park}, \bibinfo{person}{Gyeongmin Choe}, \bibinfo{person}{Jinsun Park},
  \bibinfo{person}{Yunsu Bok}, \bibinfo{person}{Yu{-}Wing Tai}, {and}
  \bibinfo{person}{In~So Kweon}.} \bibinfo{year}{2015}\natexlab{}.
\newblock \showarticletitle{{Accurate depth map estimation from a lenslet light
  field camera}}. In \bibinfo{booktitle}{\emph{{CVPR}}}.
  \bibinfo{publisher}{{IEEE} Computer Society}, \bibinfo{pages}{1547--1555}.
\newblock
\urldef\tempurl%
\url{https://doi.org/10.1109/CVPR.2015.7298762}
\showDOI{\tempurl}


\bibitem[Khor et~al\mbox{.}(2016)]%
        {paper:vr-ar-in-surgery}
\bibfield{author}{\bibinfo{person}{Wee~Sim Khor}, \bibinfo{person}{Benjamin
  Baker}, \bibinfo{person}{Kavit Amin}, \bibinfo{person}{Adrian Chan},
  \bibinfo{person}{Ketan Patel}, {and} \bibinfo{person}{Jason Wong}.}
  \bibinfo{year}{2016}\natexlab{}.
\newblock \showarticletitle{Augmented and virtual reality in surgery—the
  digital surgical environment: applications, limitations and legal pitfalls}.
\newblock \bibinfo{journal}{\emph{Annals of translational medicine}}
  \bibinfo{volume}{4}, \bibinfo{number}{23} (\bibinfo{year}{2016}).
\newblock


\bibitem[Kingma and Ba(2014)]%
        {kingma2014adam}
\bibfield{author}{\bibinfo{person}{Diederik~P Kingma} {and}
  \bibinfo{person}{Jimmy Ba}.} \bibinfo{year}{2014}\natexlab{}.
\newblock \showarticletitle{Adam: A method for stochastic optimization}.
\newblock \bibinfo{journal}{\emph{arXiv preprint arXiv:1412.6980}}
  (\bibinfo{year}{2014}).
\newblock


\bibitem[LaViola(2000)]%
        {paper:sigchi-2020-vr-sickness-symptoms-2}
\bibfield{author}{\bibinfo{person}{Joseph~J. LaViola}.}
  \bibinfo{year}{2000}\natexlab{}.
\newblock \showarticletitle{{A Discussion of Cybersickness in Virtual
  Environments}}.
\newblock \bibinfo{journal}{\emph{{ACM} {SIGCHI} Bull.}} \bibinfo{volume}{32},
  \bibinfo{number}{1} (\bibinfo{year}{2000}), \bibinfo{pages}{47--56}.
\newblock
\urldef\tempurl%
\url{https://doi.org/10.1145/333329.333344}
\showDOI{\tempurl}


\bibitem[Lee et~al\mbox{.}(2017)]%
        {paper:conf/icip/LeeJKS17}
\bibfield{author}{\bibinfo{person}{Jiyoung Lee}, \bibinfo{person}{Hyungjoo
  Jung}, \bibinfo{person}{Youngjung Kim}, {and} \bibinfo{person}{Kwanghoon
  Sohn}.} \bibinfo{year}{2017}\natexlab{}.
\newblock \showarticletitle{{Automatic 2D-to-3D conversion using multi-scale
  deep neural network}}. In \bibinfo{booktitle}{\emph{2017 {IEEE} International
  Conference on Image Processing, {ICIP} 2017, Beijing, China, September 17-20,
  2017}}. \bibinfo{publisher}{{IEEE}}, \bibinfo{pages}{730--734}.
\newblock
\urldef\tempurl%
\url{https://doi.org/10.1109/ICIP.2017.8296377}
\showDOI{\tempurl}


\bibitem[Li et~al\mbox{.}(2024)]%
        {paper:openvr-testing-infrastructure}
\bibfield{author}{\bibinfo{person}{Shuqing Li}, \bibinfo{person}{Binchang Li},
  \bibinfo{person}{Cuiyun Gao}, {and} \bibinfo{person}{Michael~R. Lyu}.}
  \bibinfo{year}{2024}\natexlab{}.
\newblock \showarticletitle{{An Interaction Simulation and Automated Testing
  Framework for Spatial Computing Extended Reality Applications}}.
\newblock  (\bibinfo{year}{2024}).
\newblock


\bibitem[Li et~al\mbox{.}(2023)]%
        {paper:vr-software-quality}
\bibfield{author}{\bibinfo{person}{Shuqing Li}, \bibinfo{person}{Lili Wei},
  \bibinfo{person}{Yepang Liu}, \bibinfo{person}{Cuiyun Gao},
  \bibinfo{person}{Shing-Chi Cheung}, {and} \bibinfo{person}{Michael~R. Lyu}.}
  \bibinfo{year}{2023}\natexlab{}.
\newblock \showarticletitle{{Towards Modeling Software Quality of Virtual
  Reality Applications from Users' Perspectives}}.
\newblock \bibinfo{journal}{\emph{arXiv preprint arXiv:2308.06783}}
  (\bibinfo{year}{2023}).
\newblock
\urldef\tempurl%
\url{https://doi.org/10.48550/ARXIV.2308.06783}
\showDOI{\tempurl}


\bibitem[Li et~al\mbox{.}(2020)]%
        {paper:issre-webxr-empirical}
\bibfield{author}{\bibinfo{person}{Shuqing Li}, \bibinfo{person}{Yechang Wu},
  \bibinfo{person}{Yi Liu}, \bibinfo{person}{Dinghua Wang},
  \bibinfo{person}{Ming Wen}, \bibinfo{person}{Yida Tao},
  \bibinfo{person}{Yulei Sui}, {and} \bibinfo{person}{Yepang Liu}.}
  \bibinfo{year}{2020}\natexlab{}.
\newblock \showarticletitle{{An Exploratory Study of Bugs in Extended Reality
  Applications on the Web}}. In \bibinfo{booktitle}{\emph{{ISSRE}}}.
  \bibinfo{publisher}{{IEEE}}, \bibinfo{pages}{172--183}.
\newblock
\urldef\tempurl%
\url{https://doi.org/10.1109/ISSRE5003.2020.00025}
\showDOI{\tempurl}


\bibitem[Liu et~al\mbox{.}(2008)]%
        {liu2008isolation}
\bibfield{author}{\bibinfo{person}{Fei~Tony Liu}, \bibinfo{person}{Kai~Ming
  Ting}, {and} \bibinfo{person}{Zhi-Hua Zhou}.}
  \bibinfo{year}{2008}\natexlab{}.
\newblock \showarticletitle{Isolation forest}. In
  \bibinfo{booktitle}{\emph{2008 eighth ieee international conference on data
  mining}}. IEEE, \bibinfo{pages}{413--422}.
\newblock


\bibitem[Liu et~al\mbox{.}(2020)]%
        {paper:owl-eyes}
\bibfield{author}{\bibinfo{person}{Zhe Liu}, \bibinfo{person}{Chunyang Chen},
  \bibinfo{person}{Junjie Wang}, \bibinfo{person}{Yuekai Huang},
  \bibinfo{person}{Jun Hu}, {and} \bibinfo{person}{Qing Wang}.}
  \bibinfo{year}{2020}\natexlab{}.
\newblock \showarticletitle{Owl Eyes: Spotting {UI} Display Issues via Visual
  Understanding}. In \bibinfo{booktitle}{\emph{{ASE}}}.
  \bibinfo{publisher}{{IEEE}}, \bibinfo{pages}{398--409}.
\newblock
\urldef\tempurl%
\url{https://doi.org/10.1145/3324884.3416547}
\showDOI{\tempurl}


\bibitem[Macklon et~al\mbox{.}(2022)]%
        {paper:html5-canvas-test}
\bibfield{author}{\bibinfo{person}{Finlay Macklon},
  \bibinfo{person}{Mohammad~Reza Taesiri}, \bibinfo{person}{Markos Viggiato},
  \bibinfo{person}{Stefan Antoszko}, \bibinfo{person}{Natalia Romanova},
  \bibinfo{person}{Dale Paas}, {and} \bibinfo{person}{Cor{-}Paul Bezemer}.}
  \bibinfo{year}{2022}\natexlab{}.
\newblock \showarticletitle{Automatically Detecting Visual Bugs in {HTML5}
  Canvas Games}. In \bibinfo{booktitle}{\emph{{ASE}}}.
  \bibinfo{publisher}{{ACM}}, \bibinfo{pages}{15:1--15:11}.
\newblock
\urldef\tempurl%
\url{https://doi.org/10.1145/3551349.3556913}
\showDOI{\tempurl}


\bibitem[McCauley and Sharkey(1992)]%
        {paper:cybersickness-name-1}
\bibfield{author}{\bibinfo{person}{Michael~E McCauley} {and}
  \bibinfo{person}{Thomas~J Sharkey}.} \bibinfo{year}{1992}\natexlab{}.
\newblock \showarticletitle{Cybersickness: Perception of self-motion in virtual
  environments}.
\newblock \bibinfo{journal}{\emph{Presence: Teleoperators \& Virtual
  Environments}} \bibinfo{volume}{1}, \bibinfo{number}{3}
  (\bibinfo{year}{1992}), \bibinfo{pages}{311--318}.
\newblock


\bibitem[Mirza and Osindero(2014)]%
        {mirza2014conditional}
\bibfield{author}{\bibinfo{person}{Mehdi Mirza} {and} \bibinfo{person}{Simon
  Osindero}.} \bibinfo{year}{2014}\natexlab{}.
\newblock \showarticletitle{Conditional generative adversarial nets}.
\newblock \bibinfo{journal}{\emph{arXiv preprint arXiv:1411.1784}}
  (\bibinfo{year}{2014}).
\newblock


\bibitem[Munafo et~al\mbox{.}(2017)]%
        {paper:vr-sickness-15-mins}
\bibfield{author}{\bibinfo{person}{Justin Munafo}, \bibinfo{person}{Meg
  Diedrick}, {and} \bibinfo{person}{Thomas~A Stoffregen}.}
  \bibinfo{year}{2017}\natexlab{}.
\newblock \showarticletitle{{The Virtual Reality Head-Mounted Display Oculus
  Rift Induces Motion Sickness and is Sexist in Its Effects}}.
\newblock \bibinfo{journal}{\emph{Experimental brain research}}
  \bibinfo{volume}{235} (\bibinfo{year}{2017}), \bibinfo{pages}{889--901}.
\newblock


\bibitem[Nichols and Patel(2002)]%
        {paper:nichols2002-vr-health-4}
\bibfield{author}{\bibinfo{person}{Sarah Nichols} {and}
  \bibinfo{person}{Harshada Patel}.} \bibinfo{year}{2002}\natexlab{}.
\newblock \showarticletitle{{Health and Safety Implications of Virtual Reality:
  A Review of Empirical Evidence}}.
\newblock \bibinfo{journal}{\emph{Applied Ergonomics}} \bibinfo{volume}{33},
  \bibinfo{number}{3} (\bibinfo{year}{2002}), \bibinfo{pages}{251--271}.
\newblock


\bibitem[Nusrat et~al\mbox{.}(2021)]%
        {paper:vr-performance-optimization}
\bibfield{author}{\bibinfo{person}{Fariha Nusrat}, \bibinfo{person}{Foyzul
  Hassan}, \bibinfo{person}{Hao Zhong}, {and} \bibinfo{person}{Xiaoyin Wang}.}
  \bibinfo{year}{2021}\natexlab{}.
\newblock \showarticletitle{{How Developers Optimize Virtual Reality
  Applications: {A} Study of Optimization Commits in Open Source Unity
  Projects}}. In \bibinfo{booktitle}{\emph{{ICSE}}}.
  \bibinfo{publisher}{{IEEE}}, \bibinfo{pages}{473--485}.
\newblock
\urldef\tempurl%
\url{https://doi.org/10.1109/ICSE43902.2021.00052}
\showDOI{\tempurl}


\bibitem[Paszke et~al\mbox{.}(2019)]%
        {paszke2019pytorch}
\bibfield{author}{\bibinfo{person}{Adam Paszke}, \bibinfo{person}{Sam Gross},
  \bibinfo{person}{Francisco Massa}, \bibinfo{person}{Adam Lerer},
  \bibinfo{person}{James Bradbury}, \bibinfo{person}{Gregory Chanan},
  \bibinfo{person}{Trevor Killeen}, \bibinfo{person}{Zeming Lin},
  \bibinfo{person}{Natalia Gimelshein}, \bibinfo{person}{Luca Antiga},
  {et~al\mbox{.}}} \bibinfo{year}{2019}\natexlab{}.
\newblock \showarticletitle{Pytorch: An imperative style, high-performance deep
  learning library}.
\newblock \bibinfo{journal}{\emph{Advances in neural information processing
  systems}}  \bibinfo{volume}{32} (\bibinfo{year}{2019}).
\newblock


\bibitem[Peng et~al\mbox{.}(2020)]%
        {paper:walkingvibe-chi20}
\bibfield{author}{\bibinfo{person}{Yi{-}Hao Peng}, \bibinfo{person}{Carolyn
  Yu}, \bibinfo{person}{Shi{-}Hong Liu}, \bibinfo{person}{Chung{-}Wei Wang},
  \bibinfo{person}{Paul Taele}, \bibinfo{person}{Neng{-}Hao Yu}, {and}
  \bibinfo{person}{Mike~Y. Chen}.} \bibinfo{year}{2020}\natexlab{}.
\newblock \showarticletitle{{WalkingVibe: Reducing Virtual Reality Sickness and
  Improving Realism while Walking in {VR} using Unobtrusive Head-mounted
  Vibrotactile Feedback}}. In \bibinfo{booktitle}{\emph{{CHI}}}.
  \bibinfo{publisher}{{ACM}}, \bibinfo{pages}{1--12}.
\newblock
\urldef\tempurl%
\url{https://doi.org/10.1145/3313831.3376847}
\showDOI{\tempurl}


\bibitem[Penn and Hout(2018)]%
        {paper:s3d-1}
\bibfield{author}{\bibinfo{person}{Rebecca~A Penn} {and}
  \bibinfo{person}{Michael~C Hout}.} \bibinfo{year}{2018}\natexlab{}.
\newblock \showarticletitle{{Making Reality Virtual: How VR "Tricks" Your
  Brain}}.
\newblock \bibinfo{journal}{\emph{{Frontiers for Young Minds}}}
  \bibinfo{volume}{6} (\bibinfo{year}{2018}).
\newblock


\bibitem[Rafi et~al\mbox{.}(2022)]%
        {paper:conf/kbse/RafiZW22}
\bibfield{author}{\bibinfo{person}{Tahmid Rafi}, \bibinfo{person}{Xueling
  Zhang}, {and} \bibinfo{person}{Xiaoyin Wang}.}
  \bibinfo{year}{2022}\natexlab{}.
\newblock \showarticletitle{PredART: Towards Automatic Oracle Prediction of
  Object Placements in Augmented Reality Testing}. In
  \bibinfo{booktitle}{\emph{37th {IEEE/ACM} International Conference on
  Automated Software Engineering, {ASE} 2022, Rochester, MI, USA, October
  10-14, 2022}}. \bibinfo{publisher}{{ACM}}, \bibinfo{pages}{77:1--77:13}.
\newblock
\urldef\tempurl%
\url{https://doi.org/10.1145/3551349.3561160}
\showDOI{\tempurl}


\bibitem[Ranftl et~al\mbox{.}(2020)]%
        {ranftl2020towards}
\bibfield{author}{\bibinfo{person}{Ren{\'e} Ranftl}, \bibinfo{person}{Katrin
  Lasinger}, \bibinfo{person}{David Hafner}, \bibinfo{person}{Konrad
  Schindler}, {and} \bibinfo{person}{Vladlen Koltun}.}
  \bibinfo{year}{2020}\natexlab{}.
\newblock \showarticletitle{Towards robust monocular depth estimation: Mixing
  datasets for zero-shot cross-dataset transfer}.
\newblock \bibinfo{journal}{\emph{IEEE transactions on pattern analysis and
  machine intelligence}} \bibinfo{volume}{44}, \bibinfo{number}{3}
  (\bibinfo{year}{2020}), \bibinfo{pages}{1623--1637}.
\newblock


\bibitem[Rodriguez and Wang(2017)]%
        {paper:os-vr-rodriguez17}
\bibfield{author}{\bibinfo{person}{Irving Rodriguez} {and}
  \bibinfo{person}{Xiaoyin Wang}.} \bibinfo{year}{2017}\natexlab{}.
\newblock \showarticletitle{{An Empirical Study of Open Source Virtual Reality
  Software Projects}}. In \bibinfo{booktitle}{\emph{{ESEM}}}.
  \bibinfo{publisher}{{IEEE} Computer Society}, \bibinfo{pages}{474--475}.
\newblock
\urldef\tempurl%
\url{https://doi.org/10.1109/ESEM.2017.65}
\showDOI{\tempurl}


\bibitem[Ronneberger et~al\mbox{.}(2015)]%
        {paper:conf/miccai/RonnebergerFB15}
\bibfield{author}{\bibinfo{person}{Olaf Ronneberger}, \bibinfo{person}{Philipp
  Fischer}, {and} \bibinfo{person}{Thomas Brox}.}
  \bibinfo{year}{2015}\natexlab{}.
\newblock \showarticletitle{U-Net: Convolutional Networks for Biomedical Image
  Segmentation}. In \bibinfo{booktitle}{\emph{Medical Image Computing and
  Computer-Assisted Intervention - {MICCAI} 2015 - 18th International
  Conference Munich, Germany, October 5 - 9, 2015, Proceedings, Part {III}}}
  \emph{(\bibinfo{series}{Lecture Notes in Computer Science},
  Vol.~\bibinfo{volume}{9351})}, \bibfield{editor}{\bibinfo{person}{Nassir
  Navab}, \bibinfo{person}{Joachim Hornegger}, \bibinfo{person}{William
  M.~Wells III}, {and} \bibinfo{person}{Alejandro~F. Frangi}} (Eds.).
  \bibinfo{publisher}{Springer}, \bibinfo{pages}{234--241}.
\newblock
\urldef\tempurl%
\url{https://doi.org/10.1007/978-3-319-24574-4\_28}
\showDOI{\tempurl}


\bibitem[Rzig et~al\mbox{.}(2023)]%
        {paper:vr-testing-empirical}
\bibfield{author}{\bibinfo{person}{Dhia~Elhaq Rzig}, \bibinfo{person}{Nafees
  Iqbal}, \bibinfo{person}{Isabella Attisano}, \bibinfo{person}{Xue Qin}, {and}
  \bibinfo{person}{Foyzul Hassan}.} \bibinfo{year}{2023}\natexlab{}.
\newblock \showarticletitle{{Virtual Reality {(VR)} Automated Testing in the
  Wild: {A} Case Study on Unity-Based {VR} Applications}}. In
  \bibinfo{booktitle}{\emph{{ISSTA}}},
  \bibfield{editor}{\bibinfo{person}{Ren{\'{e}} Just} {and}
  \bibinfo{person}{Gordon Fraser}} (Eds.). \bibinfo{publisher}{{ACM}},
  \bibinfo{pages}{1269--1281}.
\newblock
\urldef\tempurl%
\url{https://doi.org/10.1145/3597926.3598134}
\showDOI{\tempurl}


\bibitem[Shih et~al\mbox{.}(2020)]%
        {paper:conf/cvpr/ShihSKH20}
\bibfield{author}{\bibinfo{person}{Meng{-}Li Shih},
  \bibinfo{person}{Shih{-}Yang Su}, \bibinfo{person}{Johannes Kopf}, {and}
  \bibinfo{person}{Jia{-}Bin Huang}.} \bibinfo{year}{2020}\natexlab{}.
\newblock \showarticletitle{{3D Photography Using Context-Aware Layered Depth
  Inpainting}}. In \bibinfo{booktitle}{\emph{{CVPR}}}.
  \bibinfo{publisher}{Computer Vision Foundation / {IEEE}},
  \bibinfo{pages}{8025--8035}.
\newblock
\urldef\tempurl%
\url{https://doi.org/10.1109/CVPR42600.2020.00805}
\showDOI{\tempurl}


\bibitem[Souza et~al\mbox{.}(2018)]%
        {paper:journals/stvr/SouzaND18}
\bibfield{author}{\bibinfo{person}{Alinne Cristinne~Corr{\^{e}}a Souza},
  \bibinfo{person}{F{\'{a}}tima L.~S. Nunes}, {and}
  \bibinfo{person}{M{\'{a}}rcio~Eduardo Delamaro}.}
  \bibinfo{year}{2018}\natexlab{}.
\newblock \showarticletitle{{An Automated Functional Testing Approach for
  Virtual Reality Applications}}.
\newblock \bibinfo{journal}{\emph{Softw. Test. Verification Reliab.}}
  \bibinfo{volume}{28}, \bibinfo{number}{8} (\bibinfo{year}{2018}).
\newblock
\urldef\tempurl%
\url{https://doi.org/10.1002/STVR.1690}
\showDOI{\tempurl}


\bibitem[Statista(2022)]%
        {website:vr-users-number}
\bibfield{author}{\bibinfo{person}{Statista}.} \bibinfo{year}{2022}\natexlab{}.
\newblock \bibinfo{title}{{Report of Active Virtual Reality Users Worldwide}}.
\newblock
  \bibinfo{howpublished}{\url{https://www.statista.com/statistics/426469/active-virtual-reality-users-worldwide/}}.
\newblock


\bibitem[Wang(2022)]%
        {paper:conf/icse/Wang22}
\bibfield{author}{\bibinfo{person}{Xiaoyin Wang}.}
  \bibinfo{year}{2022}\natexlab{}.
\newblock \showarticletitle{{VRTest: An Extensible Framework for Automatic
  Testing of Virtual Reality Scenes}}. In \bibinfo{booktitle}{\emph{{ICSE:
  Companion Proceedings}}}. \bibinfo{publisher}{{ACM/IEEE}},
  \bibinfo{pages}{232--236}.
\newblock
\urldef\tempurl%
\url{https://doi.org/10.1145/3510454.3516870}
\showDOI{\tempurl}


\bibitem[Wang et~al\mbox{.}(2023)]%
        {paper:conf/kbse/WangRM23}
\bibfield{author}{\bibinfo{person}{Xiaoyin Wang}, \bibinfo{person}{Tahmid
  Rafi}, {and} \bibinfo{person}{Na Meng}.} \bibinfo{year}{2023}\natexlab{}.
\newblock \showarticletitle{{VRGuide: Efficient Testing of Virtual Reality
  Scenes via Dynamic Cut Coverage}}. In \bibinfo{booktitle}{\emph{{ASE}}}.
  \bibinfo{publisher}{{IEEE}}, \bibinfo{pages}{951--962}.
\newblock
\urldef\tempurl%
\url{https://doi.org/10.1109/ASE56229.2023.00197}
\showDOI{\tempurl}


\bibitem[Wickramasinghe et~al\mbox{.}(2021)]%
        {wickramasinghe2021resnet}
\bibfield{author}{\bibinfo{person}{Chathurika~S Wickramasinghe},
  \bibinfo{person}{Daniel~L Marino}, {and} \bibinfo{person}{Milos Manic}.}
  \bibinfo{year}{2021}\natexlab{}.
\newblock \showarticletitle{ResNet autoencoders for unsupervised feature
  learning from high-dimensional data: Deep models resistant to performance
  degradation}.
\newblock \bibinfo{journal}{\emph{IEEE Access}}  \bibinfo{volume}{9}
  (\bibinfo{year}{2021}), \bibinfo{pages}{40511--40520}.
\newblock


\bibitem[Xie et~al\mbox{.}(2016)]%
        {paper:conf/eccv/XieGF16}
\bibfield{author}{\bibinfo{person}{Junyuan Xie}, \bibinfo{person}{Ross~B.
  Girshick}, {and} \bibinfo{person}{Ali Farhadi}.}
  \bibinfo{year}{2016}\natexlab{}.
\newblock \showarticletitle{{Deep3D: Fully Automatic 2D-to-3D Video Conversion
  with Deep Convolutional Neural Networks}}. In
  \bibinfo{booktitle}{\emph{{ECCV}}} \emph{(\bibinfo{series}{Lecture Notes in
  Computer Science}, Vol.~\bibinfo{volume}{9908})}.
  \bibinfo{publisher}{Springer}, \bibinfo{pages}{842--857}.
\newblock
\urldef\tempurl%
\url{https://doi.org/10.1007/978-3-319-46493-0\_51}
\showDOI{\tempurl}


\bibitem[Xu et~al\mbox{.}(2015)]%
        {xu2015empirical}
\bibfield{author}{\bibinfo{person}{Bing Xu}, \bibinfo{person}{Naiyan Wang},
  \bibinfo{person}{Tianqi Chen}, {and} \bibinfo{person}{Mu Li}.}
  \bibinfo{year}{2015}\natexlab{}.
\newblock \showarticletitle{Empirical evaluation of rectified activations in
  convolutional network}.
\newblock \bibinfo{journal}{\emph{arXiv preprint arXiv:1505.00853}}
  (\bibinfo{year}{2015}).
\newblock


\bibitem[Zhao et~al\mbox{.}(2020)]%
        {paper:seenomaly}
\bibfield{author}{\bibinfo{person}{Dehai Zhao}, \bibinfo{person}{Zhenchang
  Xing}, \bibinfo{person}{Chunyang Chen}, \bibinfo{person}{Xiwei Xu},
  \bibinfo{person}{Liming Zhu}, \bibinfo{person}{Guoqiang Li}, {and}
  \bibinfo{person}{Jinshui Wang}.} \bibinfo{year}{2020}\natexlab{}.
\newblock \showarticletitle{Seenomaly: vision-based linting of {GUI} animation
  effects against design-don't guidelines}. In
  \bibinfo{booktitle}{\emph{{ICSE}}}, \bibfield{editor}{\bibinfo{person}{Gregg
  Rothermel} {and} \bibinfo{person}{Doo{-}Hwan Bae}} (Eds.).
  \bibinfo{publisher}{{ACM}}, \bibinfo{pages}{1286--1297}.
\newblock
\urldef\tempurl%
\url{https://doi.org/10.1145/3377811.3380411}
\showDOI{\tempurl}


\bibitem[Zhu et~al\mbox{.}(2017)]%
        {paper:conf/iccv/ZhuPIE17}
\bibfield{author}{\bibinfo{person}{Jun{-}Yan Zhu}, \bibinfo{person}{Taesung
  Park}, \bibinfo{person}{Phillip Isola}, {and} \bibinfo{person}{Alexei~A.
  Efros}.} \bibinfo{year}{2017}\natexlab{}.
\newblock \showarticletitle{{Unpaired Image-to-Image Translation Using
  Cycle-Consistent Adversarial Networks}}. In
  \bibinfo{booktitle}{\emph{{ICCV}}}. \bibinfo{publisher}{{IEEE} Computer
  Society}, \bibinfo{pages}{2242--2251}.
\newblock
\urldef\tempurl%
\url{https://doi.org/10.1109/ICCV.2017.244}
\showDOI{\tempurl}


\bibitem[Zhu et~al\mbox{.}(2020)]%
        {paper:conf/mipr/ZhuLW20}
\bibfield{author}{\bibinfo{person}{Mingtong Zhu}, \bibinfo{person}{Xiangning
  Liu}, {and} \bibinfo{person}{Ronggang Wang}.}
  \bibinfo{year}{2020}\natexlab{}.
\newblock \showarticletitle{{Fused Network for View Synthesis}}. In
  \bibinfo{booktitle}{\emph{3rd {IEEE} Conference on Multimedia Information
  Processing and Retrieval, {MIPR} 2020, Shenzhen, China, August 6-8, 2020}}.
  \bibinfo{publisher}{{IEEE}}, \bibinfo{pages}{303--306}.
\newblock
\urldef\tempurl%
\url{https://doi.org/10.1109/MIPR49039.2020.00069}
\showDOI{\tempurl}


\end{thebibliography}

\end{document}